\documentclass[aps,
prd,
preprint,
groupedaddress,showpacs,showkeys,12pt]{revtex4-1}
\usepackage{graphics}
\usepackage{amsmath,amssymb,amsfonts} 
\usepackage{amscd}
\usepackage{amsthm}
\def\proof{{\bf Proof}.\ }
\newtheorem{Axiom}{Axiom}

\newtheorem{Lemma}{Lemma}

\newtheorem{Theorem}{Theorem}

\newtheorem{Definition}{Definition}
\newtheorem{Remark}{Remark}

\begin{document}
%\noindent
\title{\Large {Systems of Cosserat--Zhilin in Newtonian Mechanics\vspace{3pt}% A New Architecture of Newtonian Mechanics %in the Footsteps of Truesdell to Cosserat
}%\\ ({an Idiot's Fugitive Sketch of Newtonian Mechanics%: })
 }
\author{\bf Al Cheremensky\vspace{11pt}}%
\affiliation{St Petersburg State University, Faculty of Applied Mathematics and Control Processes\\  
E--mail: {cheremensky@yahoo.com}\vspace{17pt}}

\begin{abstract} 
%The paper uses the concepts of Cosserat and Zhilin in order to define the new notion of mechanical system as the main object of Newtonian mechanics. To this end new notions of vector calculus -- sliders and screw measures (bi--measures) are applied. 
%Mechanical systems of Cosserat--Zhilin are introduced as the main object of Newtonian mechanics. 
%The paper revises the concepts of rational (Newtonian) mechanics 

Mechanical systems of Cosserat--Zhilin are introduced as the main object of rational (non--relativistic) mechanics on the base of new notions of vector calculus -- sliders and screw measures (bi--measures). \vspace{3pt}

The differential equations of motion are derived for different types of Cosserat--Zhilin systems in the case where Stocks theorem is applicable. \vspace{3pt}
%The well--known systems -- mass--points, rigid bodies, body--points, classical and polar continuous media -- are treated as examples of realizations of Cosserat--Zhilin systems.
%2-- and 3--dimensional classical and polar continua are defined , the paper represents the measure of stress with the help of 

The paper defines multiplicative groups which represent the measure of stress as a (well--defined) linear isotropic map of 
 {strain} tensor or tensor of {strain velocities} for classical and polar continua in 2-- and 3--dimensional cases.%\vskip .02in
 \vspace{3pt}

\vspace{3pt}

\end{abstract} {\maketitle}
\markboth{}{}\vskip 1in
\tableofcontents \vskip 1in\newpage
%
 %\subsection{} \medskip
\section*{\noindent Introduction}\vspace{-9pt}
{\em `The ancients considered mechanics in a twofold respect; as rational, which proceeds accurately by demonstration; and practical$\ldots$ Rational mechanics will be the science of motions resulting from any forces whatsoever, and of the forces required to produce any motions, accurately proposed and demonstrated'} \cite{Newton}.

The progress of rational mechanics (and physics of XVIII--XIX centuries) is primarily based on working out its mathematical aspects. 
In 1687, I. Newton published {\em Philosophiae Naturalis Principia Mathematica} where the leading role of mathematics in rational mechanics is directly pointed out in its title. Newton 
 deliberately almost never used mathematical analysis: use new and unusual methods would jeopardize the credibility of his results. But already %By this reason he used the geometry of Euclid as a paragon.
 in 1736, in {\em Mechanics}, L. Euler explicitly stressed that 
 `{\em full understanding mechanics can be achieved only through mathematical analysis}', thus, emphasizing that mathematics should be put at the forefront and the consideration of the physical aspects only is insufficient.

The necessity of mathematization of mechanics was marked by D'Alembert in 1743: 	{\em `Rational mechanics, like geometry, must be based upon axioms which are obviously true}' \cite{Truesdell1} 
  %: the equation of balance of momentum and equation of balanceof kinetic torque (or torque of momentum in accepted, but unsuccessful, terms).In 1776
%Lagrange continued the further mathematization of rational mechanics, but it was L. Euler who when 
%We shall call the rational non--relativistic mechanics {\em Newtonian} as it is Newtonian mechanics that has opened wide avenue for studying various mechanical and physical systems (it is possible to use other names: mechanics of Galileo, Euler or Cosserat \cite{Konoplev1999, Zhilin}).
 %the realization of a certain goal -- to create an axiomatics of mechanics and thermodynamics in such a way that it covers 
The first system of axioms in mechanics was introduced by I. Newton.  % Conversely, the geometry itself `{\em is founded in mechanical practice, and is nothing but that part of universal mechanics which accurately proposes and demonstrates the art of measuring}' \cite{Newton}. % 1997).}\end{Remark}
 Now we have many systems of axioms at hand. The present paper gives one more to represent mechanics as a mathematical science and to determine the widest class of mechanical systems covered by it (may be, it is time to introduce the term of mathematical mechanics by analogy with that of mathematical physics). \vspace{-19pt} 
\section{Fundamentals of rational mechanics}   

% \vskip .4in\hskip 1.3in 
`\emph{$\ldots $the dynamics of a continuous system must clearly include as a limiting case (corresponding to a medium of density everywhere zero except in one very small region) the mechanics of a single material particle. This at once shows that it is absolutely necessary that the postulates introduced for the mechanics of a continuous system should be brought into harmony with the modifications accepted above in the mechanics of the material particle}' \cite{Levi--Civita}.

In other words, we must consider the various branches of general mechanics from a unified point of view.%
\vspace{-9pt} 
\subsection{Primitive concepts of rational mechanics}\vspace{-9pt}
The following `\hspace{-.051cm}{\em experimental facts}' lie at the 
foundation of rational mechanics \cite{Arnold}: \vskip -.15in
%\hskip 1in \vskip -.1in
 {\begin{enumerate} \vskip -.2in \em
%\begin{minipage}[t]{110mm} %\lineskip -0.5cm \parskip -0.05cm 
 \item all the natural phenomena occur in {\em space} and {\em time}.\vspace{-5pt}
\item {Galileo's principle of relativity}: `\hspace{.051cm}{\em there exist coordinate systems (called {\em inertial}) possessing the following two properties: \vspace{-18pt}%\vskip -.4in
\begin{quote}{%\parskip -.025cm 
\vskip -.2in
\item \thinspace \thinspace \hskip -.4cm -- \thinspace \thinspace all the laws of nature at all moments of time are the same in all inertial systems; \vspace{-3pt}%\newpage
\item \thinspace \thinspace \hskip -.4cm -- \thinspace \thinspace all coordinate systems in uniform rectilinear motion with respect to an inertial one are themselves inertial'}.\vspace{-3pt}%\vskip -.2in 
\end{quote}}
%\vspace{7pt}
\item {Newton's principle of determinacy}: {\em the initial state of a mechanical system uniquely determines all of its motion}. \vspace{-5pt}
 \end{enumerate}
It is easy to see that all of the above is nothing more than constants of the science language. That is why it is necessary to clear up their mathematical essence. In particular, we must answer the following questions:\vskip -.7in
\begin{quote}{\em %\parskip -0.05cm \lineskip -0.5cm 
\vskip -.6in
\item \thinspace \thinspace \hskip -.4cm -- \thinspace \thinspace what are the laws of nature, about of which Galileo's principle says?\vspace{-9pt}
%\item \thinspace \thinspace \hskip -.4cm -- \thinspace \thinspace what do we mean under `the same', {\em i.e.}?\vspace{-3pt}
\item \thinspace \thinspace \hskip -.4cm -- \thinspace \thinspace what is the group of transformations w.r.t. of which the laws are invariant'?\vspace{-9pt}
\item \thinspace \thinspace \hskip -.4cm -- \thinspace \thinspace what do we mean under `mechanical system'?\vspace{-9pt} %
\item \thinspace \thinspace \hskip -.4cm -- \thinspace \thinspace what main types of mechanical systems do  we have?} %
\end{quote}\vspace{3pt} %\medskip
%
%the Galilean's relativity principle to the above mentioned provisions, as it is in the basis of rational mechanics, too.
 %connection must be set %, At the same time the notion of body is nothing more than a constant or concept of the science language but not definition. 
%That is why we shall introduce the notion of {\em mechanical system} as a mathematical category.

{\bf \emph{Galilean space--time structure}}. In what follows, we shall use $n$--dimensional affine space ${\mathbf A}\hspace{-.05cm}^n$ modeled on $n$--dimensional vector space ${\mathbf V}^n$, $n$--dimensional Euclidean space ${\mathbf R}^n$ (${\mathbf R}= {\mathbf R}^1$ is the set of all real numbers). %We shall assume  sets in the spaces ${\mathbf A}\hspace{-.05cm}^n$ and ${\mathbf R}^n$ to be measurable in the sense of Lebesgue.{\mathbf A}\hspace{-.05cm}^

Define the Galilean space--time structure as the quadruple ${\mathbf G} = \{{\mathbf V}^n, {\mathbf A}\hspace{-.05cm}^n, \tau, g\}$ where \cite{Arnold} \vskip -.43in
\begin{quote}{\parskip -.05cm \em \vskip -.43in
\item \thinspace \thinspace \hskip -.4cm -- \thinspace \thinspace $\tau \hspace{-.15cm}: {\mathbf V}^n \rightarrow {\mathbf R}$ is a surjective linear mapping called {\em time} one, and\vspace{-3pt}
\item \thinspace \thinspace \hskip -.4cm -- \thinspace \thinspace $g=\langle \cdot,\cdot\rangle $ is an inner product on ${\rm ker} \{\tau \}$ $(={\mathbf V}^{n-1})$.} \vspace{-3pt}
\end{quote}
The space ${\mathbf A}\hspace{-.05cm}^n$ with the Galilean space--time structure is called {\em Galilean} (note that the term `Galilean' is merely traditional and should not be regarded as an attribution to Galileo). 

%In the framework of this structure the affine space $ {\mathbf A}\hspace{-.05cm}^n$ is called {\em Galilean one}. 
Time is a linear mapping $\tau \hspace{-.15cm}: {\mathbf V}^n \rightarrow {\mathbf R}$ from the vector space of parallel 
displacements of ${\mathbf A}\hspace{-.05cm}^n$ to the real `time axis'. We shall denote the range of $ \tau$  by ${\mathbf T}\subseteq {\mathbf R}$. The time interval from event $a \in {\mathbf A}\hspace{-.05cm}^n$ to event $b \in {\mathbf A}\hspace{-.05cm}^n$ is the number $\tau (b - a)$ (it is plain that $b - a\in {\mathbf V}^n$). If $\tau (b - a)=0$, then the events $a$ and $b$ are called simultaneous. 
 
The set of events simultaneous with a given event forms ${n-1}$--dimensional affine space $ {\mathbf A}\hspace{-.05cm}^{n-1} \subset {\mathbf A}\hspace{-.05cm}^n$ modeled on ${\rm ker} \{\tau \}$. It is called {\em space of simultaneous events}. 

There is the group of affine transformations of the space $ {\mathbf A}\hspace{-.05cm}^n$ which preserve the Galilean time--space structure. The elements of this group are called {\em Galilean} transformations. They preserve intervals of time and the distance between simultaneous events \cite{Arnold}. 
\begin{Theorem} Each Galilean transformation is movement of the space of simultaneous events, accompanied by a shift of the origin of time \cite{Arnold, Golubev}. %(Arnold 1989, Golubev 2000).
\end{Theorem}
The inner product $\langle \cdot,\cdot\rangle $ (in Galilean space--time) enables one to pass from the affine space ${\mathbf A}\hspace{-.05cm}^n$ to {\em Euclidean} space ${\mathbf R}^{n}$ with the distance $\rho(x,y)={\| {x-y}\|} = \sqrt{\langle {x-y},{x-y}\rangle }$ between points $x$ and $y\in {\mathbf A}\hspace{-.05cm}^n$. % and to introduce {\em Cartesian} frame ${\mathcal E}_0$ in ${\mathbf R}^{n-1}$.

The bijective map ${\mathbf A}\hspace{-.05cm}^n\rightarrow {\mathbf R}^{n-1}\times{\mathbf T}$ is called {\em frame of reference} \cite{Arnold,Golubev} (here ${\mathbf R}^{n-1}$ is also a space of simultaneous events). 
For each frame of reference the corresponding space $ {\mathbf R}^{n-1}\times{\mathbf T}$ is Galilean. That is why we shall call any frame of reference Galilean, too.

Define {\em world--line} as a curve in ${\mathbf A}\hspace{-.05cm}^n$ whose image in $ {\mathbf R}^{n-1}\times{\mathbf T}$ %(in the frame of reference)
 associates one point $x(t) \in {\mathbf R}^{n-1}$ to each instant $t \in {\mathbf T}$. It means that a world--line does not have simultaneous points. 

A collection of non--intersectional world--lines forms {\em world--tube. }\begin{Remark}Intersections of world--lines represent collisions or the creation or destruction of bodies or elements of bodies. In specific mechanical theories such intersections are usually excluded (the principle of impenetrability) altogether or allowed as exceptional cases subject to specified conditions.\end{Remark}

Let us fix a world--tube $\tilde{\mathbf \Lambda}\subset {\mathbf A}\hspace{-.05cm}^n$ and call it {\em universe}.\begin{Remark}As well as in {\em probability theory} \cite{Kolmogorov}, any universe is separately specified for every problem under consideration.\end{Remark} %Points of $\tilde{\mathbf \Lambda}$ are called {\em world points} or {\em events}. 
A given world--tube ${\mathbf \Lambda}\subset \tilde{\mathbf \Lambda}$, the world--tube ${\mathbf \Lambda}^{\hspace {-.05cm}e}=\tilde{\mathbf \Lambda}\setminus {\mathbf \Lambda}$ is called {\em environment} of ${\mathbf \Lambda}$ in the universe. 
%Let us take ar$A$ and Any bi--measure ${\it \varPsi}(A,B)$ 

The universe $\tilde{\mathbf \Lambda}$ defines the family $\{\tilde{\Lambda}_t\subset {\mathbf R}^{n-1}$, $t \in {\mathbf T}\}$. %in  a Galilean frame of reference.
For any world--tube ${\mathbf \Lambda}\subset \tilde{\mathbf \Lambda}$ we have the family $\{{\Lambda}_t\subset \tilde{\Lambda}_t$, $t \in {\mathbf T}\}$. 

Let $\sigma_{n-1}$ be $\sigma$--algebra on the space $ {\mathbf R}^{n-1}$. We shall use the following Borel measure \vspace{-3pt}
\begin{equation}
	\mu_{n-1} (A)=\mu _{ac}(A)+\mu _{pp}(A), \quad A\in \sigma_{n-1} \label{1}%\vspace{3pt}
\end{equation}
where $\mu _{ac}(A)$ is the absolutely continuous component w.r.t. Lebesgue
measure and $\mu _{pp}(A)$ is the pure point (discrete) component presented as $\mu _{pp}(A)=\sum_k\mu _ k(x_k)$ for points ${x }_k\in A$ whose are called {\em pure}, the others being called {\em continuous}
 \cite{Reed}.%; $\sigma _{n-1}$ is $\sigma $--algebra of subsets in ${\mathbf R}^{n-1}$. 
	\begin{Remark}
In some cases where, {\em e.g.}, no increment of mass is, we shall assume that if a point of $\tilde{\Lambda}_t\subset {\mathbf R}^{n-1}$ is pure (or continuous) at some time instant $t$, all other points of the corresponding world--line ($\forall t \in {\mathbf T}$) are also pure (or continuous), too. \vspace{-3pt}	
\end{Remark}
\medskip 
	
{\bf \emph{Main axioms}}. With any point ${x} (t)\in \tilde{\Lambda}_t\in {\mathbf R}^{n-1}$ we associate the position radius--vector $ r_{x} (t)=\overrightarrow{(O,x(t))}$ w.r.t. the origin  point $O\in {\mathbf R}^{n-1}$. %Introduce the translation vector $r_{y,x}\in {\bf V}_{n-1}$ from a point $y\in {\mathbf R}_{n-1}$ to $x$.
Define the translation velocity $ v_ {x} =r_{x} ^ {{\hskip.02cm} ^ \centerdot} (t)$. 
\begin{Remark}We shall denote full derivatives by $t$ with the help of the superscript ${\hspace{.02cm }^\centerdot}{\hspace{.02cm }}$, 
{\em e.g.}, for any function $f=f(x(t),t)$ we have $f^ {\hspace{.01cm }{}^\centerdot}=\frac{\partial}{\partial t}f+({\rm div}\hspace {.02cm} f)\hspace {.02cm} x^ {\hspace {.01cm}{}^ \centerdot}(t)$.\vspace{-3pt}\end{Remark}

	Let us consider an orthonormal basis $ {\mathbf e}_ {x} $ at the point $ {x} (t)\in {\mathbf R}^{n-1}$. Rotation of the basis $ {\mathbf e}_ {x} $ can be characterized by {\em torque} ${\mu_x} $ (angular velocity -- spin, angular momentum vector of mass unit or rotation tensor, {\em etc.}). 
	\begin{Definition} %The points of ${\Lambda_t}$ and the vectors $v_x$ and $\mu_{x}$, associated with them, constitute {\em state} of a mechanical system. 
The map of ${\mathbf T}$ into the set ${\Lambda_t}$ and the vectors $v_x$ and $\mu_{x}$, associated with it, is called {\em motion}.
   {\em  Mechanical interaction} is something that generates motion.  \end{Definition}
 Hereinafter we shall use the new notions of vector calculus -- sliders and screw measures (bi--measures) (see Appendix 1). 

Introduce the measure ${\mathcal P}$ as a screw having values \vspace{-3pt}
\begin{equation}
	{\mathcal P}({\Lambda_{t}})=\int \hspace{-.05cm} \chi _{\hspace {-.05cm}_{\Lambda_{t}}} l ^{\hspace{.02cm}p_x,q_x} \mu _{n-1} (dx), \quad \Lambda_{t} \subset \tilde{\Lambda}_t,\ t\in {\mathbf T} \label{P} %\vspace{-3pt}
\end{equation}
where  $\chi _{\hspace {-.05cm}_{A}}$ is the {\em characteristic function} of $A\in \sigma _{n-1}$; 
the slider $l ^{\hspace{.02cm}p_x,q_x}$ is defined by the following relation (see also \cite{Zhilin}) \vspace{3pt}
 \begin{equation}
	\begin{pmatrix}
		{p }_x \vspace{-5pt}\\ 
		q_x
		\end{pmatrix}=
	{\mathcal \theta_x}\begin{pmatrix}
		{v }_{x}\vspace{-5pt}\\  
		\mu_{x}
		\end{pmatrix} \label{4}\vspace{9pt}   
\end{equation}
Here ${\mathcal \theta_x}$ is a non--negative defined, symmetric 2nd--order tensor. 

For any two world--tubes ${\mathbf \Lambda}$ and ${\mathbf \Lambda}^{'}$ with $\Lambda_{t}$ and $\Lambda_{t}^{'}$ in the space of of simultaneous events, respectively, 
define the skew signed field bi--measure ${\it \varPhi}$ as a screw (by each argument) having values ${\it \varPhi}(\Lambda_{t},\Lambda_{t}^{'})$.

%With a common understanding the motion of bodies is caused by their mechanical interaction which is described in the terms of mass, force and torque of force  (linear and angular momenta) as well as constraints on bodies and their parts, {\em etc.} %\medskip %According to this concept, we shall accept the following statement.\begin{Definition} The mechanical interaction is that generates motion.\end{Definition}
\begin{Axiom}There exist  measures ${\mathcal P}$ and ${\it \varPhi}$ which  describe motion.\end{Axiom}
 \begin{Definition}These measures are called {\em kinetic} and {\em dynamic} measures of motion, respectively. %Mechanical interaction is something that can be expressed by them.
	\end{Definition}
There are special frames of reference that help us  to distinguish between the  postulated  kinetic and dynamic measures of motion among various measures  and bi--measures.
\begin{Axiom}	There exists a Galilean frame of reference where the  postulated kinetic and dynamic measures of motion are connected by the following relation (see also \cite{Newton,Zhilin,Berthelot,Euler1,Euler2})%\vspace{-3pt} 
\begin{equation}
	\frac{d}{dt}{\mathcal P}({\Lambda_{t}})={\mathcal F}({\Lambda_{t}}), \quad 
			{\mathcal F}({\Lambda_{t}}) \stackrel{{\rm def}}{=}\varPhi({\Lambda_{t}},\Lambda^e_{t}), \quad \Lambda_{t} \subset \tilde{\Lambda}_t,\ t\in {\mathbf T}
	\label{2} %\vspace{5pt}
\end{equation} 
\end{Axiom}
As result we may introduce the following notions:\vspace{-3pt} 
%\begin{Definition}\vskip -.1in
\begin{enumerate}%
{\parskip -.2cm \em 
\item the above mentioned frame of reference is called {\em inertial};\vspace{3pt} 
%\item ${\mathcal P}$	is called {\em kinetic measure} of motion;
\item the aggregate $\alpha = \{{\Lambda_t}\subset \tilde{\Lambda}_t ,\mu_{n-1}, {\mathcal \theta_x}, x\in {\Lambda_t},{\mathcal P},{\mathcal F}, \forall t \in {\mathbf T}\}$ is called {\em mechanical system {of Cosserat--Zhilin}}\vspace{3pt};
\item ${\mathcal F}$ is called {\em measure of impressed action} of the environment ${\mathbf \Lambda}^{\hspace {-.05cm}e}$ on ${\mathbf \Lambda}$ or {\em force} ({\em according to Glossary, Earth Observatory, NASA}: force is any external agent that causes a change in the motion of a mechanical system, or that causes stress in a fixed mechanical system);% impressed at the body;
\vspace{3pt} % and any immobile coordinate frame in ${\mathbf R}^{n-1}$ is called {\em inertial}, too;
%\item ${\it \varPhi}$ is called {\em measure of impressed interaction} between mechanical systems;
 %\item the map ${\mathbf T}\rightarrow \{{\Lambda}_{t}$, $t \in {\mathbf T} \}$ is called {\em motion};
%\item the points of ${\Lambda_t}$ and the set $H_x\in {\mathbf H}_x, x\in {\Lambda_t},$ of sliders, associated with them, (which are defined by relation (\ref{2})) define {\em state} of the world--tube ${\mathbf \Lambda}$ in ${\mathbf R}^{n-1}\times {\mathbf H}_x$;
\item the points of ${\Lambda_t}$ and the vectors $v_x$ and $\mu_{x}$, associated with them, constitute {\em state} of a mechanical system;\vspace{3pt} 
\item 	relation (\ref{2}) is called {\em motion equation};\vspace{3pt} %-- {Arnold} 1989). ;
 \item the set $\Lambda_{t}$ is called {\em (actual) shape} undergone by the mechanical system at $t \in {\mathbf T}$.}\vspace{5pt} 
 %\item ${\mathcal Q}= \frac{d}{dt}{\mathcal P}$ is called {\em dynamic measure of motion}.}
\end{enumerate}% \newpage \end{Definition}\begin{Remark}According to Glossary, Earth Observatory, NASA: force is any external agent that causes a change in the motion of a mechanical system, or that causes stress in a fixed mechanical system.\end{Remark}
Hereinafter we assume that frames of reference are inertial.
\begin{Remark}Introducing the kinetic and dynamic measures as screws we follow L. Euler who has opened a new era in developing of Newtonian mechanics: two independent Laws of Dynamics are stated for the first time in `{\rm New method of determination of motion of rigid bodies}' \cite{Euler1}.\end{Remark}
%{\bf Axiom 1.} {\em In the universe, there exists so called mechanical interaction between world--tubes.
 %The measures ${\mathcal P}$ and ${\it \varPhi}$ are called, respectively, the kinetic measure of inertia (`the quantity of motion is the measure of the same, arising from the velocity and quantity of matter conjunctly') and dynamic measure (`an impressed force is an action exerted upon a body, in order to change its state, either of rest, or of moving uniformly forward in a right line').}\vspace{3pt}% in the Galilean frame of reference \medskip 
%`{\em An essential feature of classical mechanics is the existence of special frames in which the relation between forces and the motions they produce is especially simple}' \cite{Truesdell}. The axiom does not explain how {\em a priori} kinetic and dynamic measures may be defined. That is why it is important to give 
%It is possible to define different measures and bi--measures for world--tubes but 
\begin{Axiom}The tensor $\theta_x$ and the measure $\varPhi$ do not dependent on frames of reference (see also \cite{Truesdell}).	
 \vspace{-3pt} \end{Axiom}
\begin{Remark}
	The tensor ${\mathcal \theta_x}$ defines {\em tensor measure of inertia} $\Theta$ having values \vspace{-3pt} 
			\[
				\Theta({\Lambda}_{t})=\int \hspace{-.05cm} \chi _{\hspace {-.05cm}_{{\Lambda}_{t}}}\theta_x\hspace{.01cm}\mu_{n-1} (dx), \quad \Lambda_{t} \subset \tilde{\Lambda}_t,\ t\in {\mathbf T} \vspace{-1pt} 
	\] 
	and {\em measure of kinetic energy}  (scalar measure of  motion)  ${\mathcal T}$ having values (see also \cite{soro})\vspace{5pt}
 \[			 {\mathcal T}({\Lambda}_{t})=\int \hspace{-.1cm} \chi _{\hspace {-.05cm}_{{\Lambda}_t}} 
	\begin{pmatrix}
	{v }_x \vspace{-5pt}\\ 
	\mu_x
	\end{pmatrix}^{\hspace{-.1cm}_T }\hspace{-.1cm}\theta_x
	\begin{pmatrix}
	{v }_x \vspace{-5pt}\\ 
	\mu_x
	\end{pmatrix}\mu_{n-1} (dx), \quad \Lambda_{t} \subset \tilde{\Lambda}_t,\ t\in {\mathbf T} \vspace{17pt}
\]
\end{Remark}
%\begin{Remark}It is motion equation that constitutes definition of each mechanical system  \cite{Arnold}.\end{Remark}
\begin{Remark}
In the rational mechanics, the views of Aristotle  were dominant  for over two millennia, as long as Galileo did not introduce his principle of inertia:\vspace{-3pt}
\begin{quote}%\vskip -.1in
`{\em any isolated (lonely in the world) material point preserves its present state, whether it be of rest or of moving uniformly forward in a straight line in the absolute space'}. \vskip -.2in\end{quote} 

Since it is impossible to determine the motion of an isolated point (body) relative to the absolute space, we could use the concept of reference frame as a reference body, equipped with a clock: (see, {\em e.g.}, \cite{Zhilin}): 
\begin{quote}\vspace{-3pt}
`{\em A body of reference with respect to which trajectories of an isolated (lonely in the world) particle are straightforward or a point, called inertial reference one'}.\vspace{-1pt}\end{quote}%"Many bodies of inertial reference frame is called an absolute space."

At the same time, forgetting how it is possible to talk about the single particle in the world, when another body -- the body of reference -- is entered.

But the trouble does not come alone: straight lines are passed in straight lines under an arbitrary affine transformation, {\em  i.e.}, according to the above definition, new reference frames  will be also inertial reference ones. If some new frame is accepted as the original one, we see that Newton's Second Law can not be executed in this frame of reference. 	Thus such a classical definition of  reference frames as well as Newton's First Law proves to be unsatisfactory  (see also \cite{27}).
\vspace{-9pt}
\end{Remark}
% (note that the measure ${\it \varPhi}$ is frame--independence by definition).\vspace{-3pt}Definition
%{\bf Axiom 2.} {\em The generalized Galilean group exists.}\vspace{3pt}
\subsection{The law of universal gravity}\vspace{-7pt}
 If the first two laws of motion, Newton had predecessors (the authorship of their own, he did not claim -- see page 71 (p. 50) in the Russian translation of Newton's Principia). The third law is wholly owned by Newton (predecessors to date nobody has been able to specify). Without it, there would be neither the equation of rigid body motion nor the law of universal gravity (see page 13 in the Russian translation of Newton's Principia):\vspace{-3pt}
\begin{quote}{\em `A particle attracts every other particle in the universe using a force that is directly proportional to the product of their masses and inversely proportional to the square of the distance between them'.} \end{quote}
Below we shall consider mechanical systems where \vspace{3pt}
\begin{equation}
		\mathcal \theta_x
		\stackrel{{\rm def}}{=}\rho_x\begin{bmatrix}
		I & A\vspace{-3pt} \\ 
		A^T& B
		\end{bmatrix}\label{mass}\vspace{3pt}
\end{equation}
Here $I$ is the unit tensor; $A$ and $B$ are 2nd--order tensors; $\rho_x$ is a non--negative function for any $x\in \Lambda_{t} \subset \tilde{\Lambda}_t,\ t\in {\mathbf T} $.
	
The function $\rho_x$ generates the measure ${\mathcal M}$, %being invariant w.r.t. the generalized Galilean group and 
having the values \vspace{-3pt}%\vspace{3pt}
 \[
			 {\mathcal M}({\Lambda}_{t})=\int \hspace{-.05cm} \chi _{\hspace {-.05cm}_{{\Lambda}_{t}}}\rho_x\hspace{.01cm}\mu_{n-1} (dx), \quad \Lambda_{t} \subset \tilde{\Lambda}_t,\ t\in {\mathbf T} \vspace{3pt}
 \]
The measure ${\mathcal M}$ is called {\em mass}. The set $\tilde{\Lambda}^{c}_t \subset \tilde{\Lambda}_t$ is called {\em set of concentration} of the measure ${\mathcal M}$ if ${\mathcal M}(\Lambda_t)=0$ for any set $\Lambda_t \subset \tilde{\Lambda}_t\setminus \tilde{\Lambda}^{c}_t$. We shall assume that relation (\ref{2}) is true only for ${\Lambda_t}\subset \tilde{\Lambda}^{c}_t$, $ t \in {\mathbf T}$.
 
The measure ${\mathcal M}$ is introduced as an integral defined over {actual shapes} $\Lambda _{t}$ undergone by a mechanical system. That is why there is the following relation (see Appendix 2) \vspace{-1pt} 
\[
\frac{d}{dt}{\mathcal M}({\Lambda_{t}})=\int \hspace{-.1cm} \chi _{\hspace {-.05cm}_{{\Lambda}_{t}^{ac}}}[\frac{d}{dt}\hspace {.05cm} \rho_x+({\rm div}\hskip.05cm {v_x})\rho_x]\hskip.05cm \mu_{ac}(dx)+\sum_k (\frac{d}{dt}\hspace {.05cm} \rho_{k})\hskip.05cm \mu_{pp}(x_k)%=\Delta {\mathcal M}({\Lambda_{t}})
, \quad \Lambda_{t} \subset \tilde{\Lambda}_t,\ t\in {\mathbf T}\vspace{3pt}  \]
	We shall assume that the function $\rho _x\hskip -.1cm =\hskip -.1cm \rho (x, t)$ is defined by 
the {\em continuity equation} %for the measure ${\mathcal M}$  
(see also Appendix 2) \vspace{-3pt} 
\[
\frac{d}{d t}\rho_x+({\rm div}\hskip.05cm {v_x})\rho_x=\frac {\partial}{\partial t}\rho_x+{\rm div}\hskip.05cm ({v_x}\rho_x)=
\nu _x \vspace{3pt}
\]
in continuous points and 
$
\frac{d}{dt}\rho _{k}=\nu _{k}
$ 
in pure points (here %$\nu _x$ depicts the density of gain or loss for the measure ${\mathcal M}$ per unit volume and unit time:
we may referred terms such as generation ($\nu _x> 0$) or re--movement ($\nu _x< 0$) to `sources' and `sinks', respectively).

Then \vspace{-7pt}\[ 
\frac{d}{dt}{\mathcal M}({\Lambda_{t}})\hskip -.05cm =\Delta {\mathcal M}({\Lambda_{t}}) \stackrel{{\rm def}}{=} \hskip -.1cm \int \hspace{-.1cm} \chi _{\hspace {-.05cm}_{{\Lambda}_{t}^{ac}}}\nu _x\hskip.05cm \mu_{ac}(dx)+\sum_k \nu _{k}\hspace {.05cm}\mu_{pp}(x_k)%=\Delta {\mathcal M}({\Lambda_{t}})
, \ \Lambda_{t} \subset \tilde{\Lambda}_t, t\in {\mathbf T}\vspace{3pt}  \]
We shall give a screw version of the law of {universal gravity}.  To this end let us define the measure $\varGamma$, having the values\vspace{-3pt}
 \[\varGamma({\Lambda}_{t}, {\Lambda}_{t}^{e}) =\int \hspace{-.1cm} \chi _{{\Lambda}_{t}}l^{{\gamma}_x} \rho_x \mu_{n-1}(dx),\quad {\gamma_x}=\gamma 
\int \hspace{-.1cm} \chi _{\hspace {-.05cm}_{{\Lambda}_{t}^{e}}}
r_{x,y}\hspace{.1cm}
\frac{\rho_y \mu_{n-1}(dy)}{\Vert
r_{x,y}\Vert ^{3}} \vspace{3pt} \]
on the sets ${\Lambda}_{t}\in \sigma_{n-1}$. Here $\gamma $ is the positive constant,  $r_{x,y}\in {\bf V}^{n-1}$ is the translation vector from a point $x$ to $y\in {\mathbf R}^{n-1}$.\vspace{3pt}
 \begin{Definition}The universal gravity is that is expressed by the measure ${\mathcal F}$ such that ${\mathcal F}({\Lambda}_{t})\stackrel{{\rm def}}{=} \varGamma({\Lambda}_{t}, {\Lambda}_{t}^{e})$. The screw measure ${\mathcal F}$ is called {\em measure of graviting action} of $\alpha^e$ upon $\alpha$ \cite{Konoplev1999}.\end{Definition}
 \begin{Definition}In the case of relation (\ref{mass}), the mechanical interaction is that the universal gravity generates. \end{Definition}
The concept of  universal gravity is the great intellectual achievement that Newton represented in the most outstanding book in the history of science: 
{\em  Philosophiae Naturalis Principia Mathematica} or, in modern language, {\em Mathematical Foundations of Physics}.
%Newton's Principia formulated the laws of motion and universal gravity, which dominated scientists' view of the physical universe for the next three centuries.
 By deriving Kepler's laws of planetary motion from his mathematical description of gravity, and then using the same principles to account for unknown before hyperbolic and parabolic orbits of celestial bodies, the tides, the precession of the equinoxes, and other phenomena, Newton demonstrated that the motion of objects on Earth and of celestial bodies could be described by the same principles.

It is paradoxical, and even insulting to anyone who is familiar with the revolution produced by Newton in  science, that, in the textbooks of theoretical mechanics, the law of universal gravity is not regarded. Sporadically, a particular case of the law, as the inverse square law, is derived from Kepler's three laws.  \begin{Remark}From the above follows that the inertia appearance is due not to ‘{\em inborn force of the matter, included in itself}’ but to mechanical systems belonging to the universe (\cite{Mach}).
\end{Remark}
\vspace{-9pt} 
 
\subsection{Concept of body} \vspace{-9pt}
 We shall use the following convention:\vspace{3pt}
%\vspace{-7pt} %in the case of a given mechanical system $\alpha = \{{\Lambda_t}\subset \tilde{\Lambda}_t ,\mu_{3}, {\mathcal \theta_x}, x\in {\Lambda_t}, {\mathcal F},\forall t \in {\mathbf T} \}$:\vskip -.9in
\begin{quote}{\em \parskip -0.25cm %\vskip -.2in
{\em body} is that takes some shapes ${\Lambda_t}\subset \tilde{\Lambda}_t$ in the space at some instants of time ({\em cf.} `{\rm every sensible body is in place}' -- Aristotle, Physics, III, 4, 208b27)}.\vspace{-7pt}
\end{quote}
The concept of body is the subject of various formalizations. For example, one may represent a body as a point--wise set, a differentiable manifold, a topological or measure space \cite{Konoplev1999,Pobedria} where a map into the space of shapes is considered.\vspace{-3pt}
\begin{Remark}These definitions follow from Plato's idea on the existence of two worlds: the world of ideas (eidos) and the world of things, or forms. And then we have only  `photo' of a body at each time    while the body itself is out of Plato's cave.\vspace{-5pt}
\end{Remark}
But there is a small obstacle: we must also transfer masses and forces to body shapes. 
If we do it in some way then the construction -- body with mass and force -- loses the primitive nature. In order to work out a mathematical theory we have all the necessary: shapes with kinematic, kinetic and dynamic structures attributed by them.  While the concept of a mechanical system has strict mathematical sense, the concept of body has only descriptive character, being a tribute of the very seminal tradition.
\vspace{-7pt}% \newpage
 \subsection{Generalization of the mechanical system concept}\vspace{-7pt}
The non--trivial nature of the mechanical system concept can be seen from the fact that we may postulate the following relation (see also \cite{Zhilin})\vspace{-5pt}
\begin{equation}
	\frac{d}{dt}{\mathcal P}({\Lambda_{t}})={\mathcal F}({\Lambda_{t}})+\Delta{\mathcal P}({\Lambda_{t}})+ {\mathcal R}({\Lambda_{t}}), \quad \Lambda_{t} \subset \tilde{\Lambda}_t,\ t\in {\mathbf T} 
	\label{3}	 \vspace{-7pt}
\end{equation}
where the signed field measure $\Delta{\mathcal P}$ is so called {\em increment velocity} of the measure ${\mathcal P}$, 
the  signed field measure ${\mathcal R}$ is so called {\em constraint action}.
%Note that the base point in mechanics is that the motion of bodies is caused by interaction with their environment in the universe $\tilde{\mathbf \Lambda}$ which is described in the terms of force, moment of force and intrinsic torque \cite{Vallander}. However one must pay attention to constraints on bodies and their parts as well as interchange of masses, linear and angular momenta. 

Below we shall assume that the measure ${\mathcal R}$ is 
formed by internal and external constraints: ${\mathcal R}={\mathcal R}_{int}+{\mathcal R}_{ext}$. Here the signed field measures ${\mathcal R}_{int}$ and ${\mathcal R}_{ext}$ have the values ${\mathcal R}_{int}({\Lambda_{t}})$ and ${\mathcal R}_{ext}({\Lambda_{t}})$ on the sets ${\Lambda}_{t}$, respectively.\vskip -.9in

 \section{Implementation of the axioms on examples of main types of Cosserat--Zhilin  systems}\vspace{-7pt}%\vskip .1in%\vspace{11pt}
\begin{quote}{\parskip -0.25cm \em \vskip .132in
`The goal of Newton was to give an answer to the question whether there is a simple rule for calculating the total movement of the heavenly bodies of our planetary system at a given state of motion of all the bodies in a given time? From observations of Tycho Brahe, Kepler deduced empirical laws of planetary motion but they were demanded an explanation. Today, everyone knows what a great, truly bee, hard work has been required to establish these laws, on the basis of empirically determined orbits. But few who imagines the genius of the method by which Kepler has defined the true orbit, based on the apparent, {\em i.e.}, of the observed motions of the Earth. These laws provide a complete description of the motion of the planets around the Sun: elliptical orbits, equality sectorial velocity ratio between the semi-major axes and periods of treatment. But these laws do not satisfy the requirement of a causal explanation. They were the three logically independent of each other rules deprived of any internal connection. The third law cannot be quantified unequivocally transferred to another, other than the Sun, the central body (there is, for example, no connection between the orbital period of the planet around the Sun and the orbital period of the satellite around its planet). But the important thing is that the laws of motion are generally not possible to derive from the state of motion at some point in time a different state in time immediately following the first. In modern terminology, we would say that they are integral laws and not differential.\vspace{11pt}

Differential law is the sole form of causal explanation, which can fully meet modern physics%mateA clear understanding of the differential law is one of the greatest intellectual achievements of Newton
'}  \cite{Einstein}.\end{quote}%%\subsection{Cosserat--Zhilin systems
\vspace{-9pt}}
Below, we are about to show the implementation of the axioms on examples of main types of Cosserat--Zhilin  systems and derive their equations of motion in the case where Stocks theorem is applicable (see \cite{Pobedria}). These equations are invariants of the generalized Galilean group (see Appendix 3 and \cite{Konoplev1999}).

 In mechanics, the most important cases of motion are in one--, two-- and  three--dimensional spaces. For the sake of brevity, hereinafter we shall only consider  \underline{the three--dimensional case}.\vspace{-7pt} %\newpage
\subsection{A body--point} \vspace{-7pt} 

Let the image of a world--line ${\mathbf \Lambda}\subset \tilde{\mathbf \Lambda}$ be the curve $\{{x}(t)\in {\Lambda}_t, t \in {\mathbf T}\}$ in ${\mathbf R}^{n-1}\times {\mathbf T}$. Assume that the points $x$ of $x(t)$ are pure. Then the corresponding mechanical system is called {\em body--point}.
\begin{Remark}This concept is not the same as in \cite{Truesdell} (see also \cite{Zhilin}).\end{Remark}
\begin{Remark}In physics, if a body has an infinitely small size and a finite mass, it is called a {\em material point} or {\em  mass--point}. We could put an electron on the role of material point because its size is extremely small, and it has some mass. However, motion of electrons can be not only translational, but also rotational. The latter does not meet the concept of material point (its rotation is not defined). Thus, {\em a priori}, we cannot consider a body with an infinitely small size and a finite mass as a material point. That is why we hope that the motion of an electron may  be described as that of a body--point with finite mass and charge. The free motion of such a mechanical system may not be uniform and rectilinear \cite{Zhilin}.  \vspace{-3pt}
\end{Remark}
 For body--points, in the case (\ref{mass}), from relations (\ref{4}) and (\ref{3}) follows that\vspace{5pt}%\vspace{-3pt} %\vspace{7pt} 
\begin{equation} 
	\frac{d}{dt}\rho_{x} l^{{\tilde p}_{x},{\tilde q}_{x}}=l^{\alpha_{x},\beta_{x}}, \quad \begin{pmatrix}
	{\tilde p}_{x}\vspace{-5pt}\\  
	{\tilde q}_{x}
	\end{pmatrix}= \begin{bmatrix}
	I & A \vspace{-5pt}\\ 
		A^T& B
		\end{bmatrix}\hspace{-.05cm}\begin{pmatrix}
	{v }_{x}\vspace{-5pt}\\  
	\mu_{x}
	\end{pmatrix}, \quad x\in x(t) \subset {\Lambda}_t,\ t\in {\mathbf T}\label{5}\vspace{3pt}%\vspace{9pt}
\end{equation} or (see also Appendix 2)\vspace{5pt}% \vspace{-3pt}
 \[ (\rho_x
		\frac{d}{dt} 
			+\nu _x) \hskip.05cm \begin{bmatrix}
	I & A \vspace{-5pt}\\ 
		A^T& B
	\end{bmatrix}
		\begin{pmatrix}
		{v_x} \vspace{-5pt}\\ 
		{\mu_x}
		\end{pmatrix}%+\begin{pmatrix}		A{\mu_x}\times {v_x} \vspace{-5pt}\\ 		B{\mu_x}\times {v_x}		\end{pmatrix}
     =\begin{pmatrix}\alpha_{x}
	\vspace{-5pt}\\ 
		\beta_{x}
		\end{pmatrix}, \quad x\in x(t) \subset {\Lambda}_t,\ t\in {\mathbf T} \vspace{9pt}
\]
\subsection{A polar medium} \vspace{-7pt}

We shall assume that all points of ${\Lambda_t}\subset \tilde{\Lambda}_t$ are continuous and $ {\Lambda_t} $ has the surface $\partial \Lambda_t$ which is Lyapunov's simple closed one (see also \cite{Zhilin}).

Introduce so called measure ${\mathcal F}$ of mass action, having the values \vspace{-3pt}
\[
	{\mathcal F}({\Lambda}_{t})=
	\int \hspace{-.1cm} \chi _{\hspace {-.05cm}_{{\Lambda}_{t}}}\rho_x l^{\hskip.02cm \gamma_x ,\delta_x, wr}\mu_3(dx), \quad \Lambda_{t} \subset \tilde{\Lambda}_t,\ t\in {\mathbf T}  
\]
 % We shall assume that all points of ${\Lambda_t}\subset \tilde{\Lambda}_t$ are continuous. 
Let constraints being in a small vicinity of $ x\in {\Lambda_{t}}$ cause the measure of {\em stress}having the values  \cite{Kilchevsky} %\vspace{3pt}
\[
	{\mathcal R}_{int}({\Lambda}_{t})=
	\int \hspace{-.1cm} \chi _{\hspace {-.05cm}_{\partial{\Lambda_{t}}}}l^{\hskip.02cm{P}_x, Q_x, wr}n_x\mu_2(dx), \quad \Lambda_{t} \subset \tilde{\Lambda}_t,\ t\in {\mathbf T}%\vspace{-3pt} 
\]
%on the sets ${\Lambda}_{t}$. 
Hereinafter $\mu_2$ is the restriction of $\mu_3$ on the surface ${\partial{\Lambda_{t}}}$, $ {n_x}$ is the normal to this surface; ${P}_x$ and ${Q}_x$ are 2nd--order tensors.
\begin{Definition}The mechanical system 
	 $\alpha = \{{\Lambda_t}\subset \tilde{\Lambda}_t ,\mu_3, \nu _x,{\mathcal \theta_x},{\mathcal F}, {\mathcal R}_{ext}, {\mathcal R}_{int}, \forall t \in {\mathbf T}\}$ is called {\em polar medium} \cite{Cosserat}. \vspace{-3pt}
\end{Definition}
For the sake of brevity assume that ${\mathcal R}_{ext}\equiv 0$. 

The following statement will be used below.
\begin{Lemma}\label{Lemma 1}Let the tensor--function ${P}_x$ be continuously differentiable in the fit region ${\Lambda_{t}}$. 
Then \cite{Vallander}\vspace{-7pt}
 \[\int \hspace{-.1cm} \chi _{\hspace {-.05cm}_{{\Lambda_{t}}}}
		{\rm div} (R_{y,x} {\hspace{.02cm}{P}_x})		\mu_3(dx)=
		\int \hspace{-.1cm} \chi _{\hspace {-.05cm}_{{\Lambda_{t}}}}
	(R_{y,x} {\hspace{.02cm}	{\rm div} {P}_x}+{\tau _x})		\mu_3(dx), \quad \Lambda_{t} \subset \tilde{\Lambda}_t,\ t\in {\mathbf T} % \vspace{3pt}
		\] 
 where the skew 2nd--order tensor $R_{y,x}$ is generated by the translation vector ${r_{y,x}}$: $R_{y,x}=r_{y,x}^\times$; ${\tau _x}$ is the dual vector to ${P}_x^T-{P}_x$ (see Appendix 1). 
\vspace{-3pt}\end{Lemma}
In the case where Stocks theorem is applicable, due to lemma \ref{Lemma 1}, from relation (\ref{3}) follows (see Appendix 2)  \vspace{3pt}
\[(\rho_x
		\frac{d}{dt} 
			+\nu _x) \hskip.05cm \begin{bmatrix}
	I & A \vspace{-5pt}\\ 
		A^T& B
	\end{bmatrix}
		\begin{pmatrix}
		{v_x} \vspace{-5pt}\\ 
		{\mu_x}
		\end{pmatrix}%+\begin{pmatrix}		A{\mu_x}\times {v_x} \vspace{-5pt}\\ 		B{\mu_x}\times {v_x}		\end{pmatrix}
		=\rho_x\begin{pmatrix}
		{\gamma_x}\vspace{-5pt}\\ 
		{\delta_x}
		\end{pmatrix} 
		+{\rm div}\begin{pmatrix}
		{P}_x\vspace{-5pt}\\ 
		{Q}_x
		\end{pmatrix}\hskip.05cm +\begin{pmatrix}
			o\vspace{-5pt}\\ 
		{\tau _x}
		\end{pmatrix} \vspace{7pt}
\]
where $o$ is the null vector.

Take a point $y(t)$ in a small vicinity of $x(t)\in {\Lambda}_{t}$ at an instant $t\in {\mathbf T}$ and define their radius--vectors $ {r_x}$ and $ {r_y}$ and the vector $ h (t)= {r_y}- {r_x}$. Then there is the following relation \vspace{-3pt} 
 \[
{v_y}(t)\cong {v_x}(t)+d{v_x}/d{r_x}h(t)\vspace{-3pt}\]
Define the tensor ${S}_x(t)$ as the solution of the following equation \vspace{-3pt}
\[
	{S}_x^{\hspace{.02cm }{}^\centerdot}(t)=d {v_x}/d{r_x} \vspace{-3pt}
\]
where its initial data are defined by so called {\em deformed} state of the medium.
\begin{Definition}${S}_x$ and ${S}_x^{\hspace{.02cm }{}^\centerdot}$ are called {\em strain} tensor and tensor of {\em strain velocities} at the point $x\in {\Lambda}_t$ at the instant $t$, respectively \cite{Konoplev1999}.
\end{Definition}
Denote the tensor ${S}_x$ or ${S}_x^{\hspace{.02cm }{}^\centerdot}$ as ${Z}_x$.
\begin{Definition} The polar medium is called {\em that of Hooke class} if the tensors ${{P}_x}$ and ${Q}_x$ are linear {\em isotropic} maps of ${Z}_x$, {\em i.e.}, invariant w.r.t. {Galilean group} of transformations.
\end{Definition}
\begin{Remark}
If ${Z}_x={S}_x$ the medium is called {\em elastic material}, if ${Z}_x={S}_x^{\hspace{.02cm }{}^\centerdot}$ the medium
is called {\em viscous fluid} \cite{Lur'e}.	
\end{Remark}
 \begin{Definition}
We shall call an isotropic matrix function of entries of $Z$ {\em well--defined} if it is invertible
  (see also \cite{Konoplev1999}). \end{Definition}
The set of invertible linear isotropic matrix functions forms a multiplicative group (see Appendix 4). 

Regarding the implementation of polar media -- see, {\em e.g.}, \cite{Pobedria, Vallander,Nowacki} and Appendix 5.
\vspace{-9pt}
\subsection{A mass--point}\vspace{-7pt}
 
In what follows, we shall assume that $A$ and $B$ are zero in relation (\ref{mass}), {\em i.e.}, we shall use homogeneous sliders and the measure ${\mathcal P}$ having the values\vspace{-3pt}
\[{\mathcal P}({\Lambda_{t}})=\int \hspace{-.1cm} \chi _{\hspace {-.05cm}_{\Lambda_{t}}} \rho_x l ^{\hspace{.02cm}v_x} \mu _3 (dx), \quad \Lambda_{t} \subset \tilde{\Lambda}_t, t\in {\mathbf T} % \vspace{3pt}	
 \] 
Assume that the increment velocity of ${\mathcal P}$ is the measure $\Delta{\mathcal P}$ having the values \vspace{-3pt}
\begin{equation}
	\Delta{\mathcal P}({\Lambda}_{t})=\int \hspace{-.1cm} \chi _{\hspace {-.05cm}_{{\Lambda}_{t}}} l^{\hspace{.02cm}\xi_x}\hspace{.005cm} \mu_3 (dx), \quad \Lambda_{t} \subset \tilde{\Lambda}_t, t\in {\mathbf T} \label{7}\vspace{-7pt}
\end{equation}
where the slider $l^{\hspace{.02cm}\xi _x}$ is its density.

Consider a world--line ${\mathbf \Lambda}\subset \tilde{\mathbf \Lambda}$ whose image in ${\mathbf R}^{3}\times {\mathbf T}$ generates the curve $\{{x }(t)\in {\Lambda}_t, t \in {\mathbf T}\}$. Assume that the points $x(t)$ are pure.
 
 Let $ {f_x}$ be the impressed force acting at the point $x=x_k\in {\Lambda}_t$ with the mass ${\mathcal M}=\rho _x \mu _{pp}$. Then the mechanical system $\alpha = \{x(t)\in {\Lambda}_t, \mu _{pp}, {f_x},\nu _x,\rho_x, {\xi_x},\forall t \in {\mathbf T}\}$ is called {\em mass--point}. 

From relation (\ref{3}) follows that \vspace{-7pt}
\begin{equation}(\rho _x 
\frac{d}{dt} +\nu _x ){v_x}= {f_x}+ {\xi _x }
\label{8}\vspace{-1pt}
\end{equation}
If $\nu _x\equiv0$ and $ {\xi _x} \equiv0$, then equation (\ref{8})
is known as {\em Newton's Second Law}. 

If $\nu _x\neq 0$ and $ {\xi _x}=\nu _x {u_x}$ where $ {u_x}$ is the velocity of mass gain or loss, then equation 
(\ref{8}) is known as that of Meshchersky \cite{27}. 
\begin{Remark}A classical example of mass--points with constraints is the mechanical system known as pendulum. \vspace{-9pt}
\end{Remark} 
\subsection{A rigid body}\vspace{-7pt}% {Hilbert} 1900,
`{\em One might try to derive the laws of the motion of rigid bodies by a limiting process from a system of axioms depending upon the idea of continuously varying conditions of a material filling all space continuously}' \cite{Hilbert}. Let us do it.

For the sake of brevity assume that ${\Delta\mathcal P}$ and ${\mathcal R}_{ext}\equiv 0$. 

The mechanical system 
$\alpha = \{{\Lambda_t}\subset \tilde{\Lambda}_t,\mu_3, \rho_x, {\mathcal F}, {\mathcal R}_{int} ,\forall t \in {\mathbf T}\}$ 
 is called {\em rigid body} if \vspace{9pt}
\begin{quote}{\em \parskip -0.25cm \lineskip -0.5cm \vskip -.3in
\item \thinspace \thinspace \hskip -.4cm -- \thinspace \thinspace the sets $\Lambda_t$ are bounded and closed;%\vspace{-3pt}
\item \thinspace \thinspace \hskip -.4cm -- \thinspace \thinspace the constraints applied on its points keep distances between them not changing with time;
\item \thinspace \thinspace \hskip -.4cm -- \thinspace \thinspace the internal constraints are {\em ideal}} \cite{Vilke}.\vspace{-3pt}
\end{quote}
A rigid body may comprise continuous and pure points.

Below we are going to obtain its motion equation with using so called {\em quasi--velocities} (Newton--Euler equations in quasi--velocities), some parametrizations of rotation matrices and {\em generalized coordinates and velocities} (Lagrange equation of II kind).\medskip 

{\bf \emph{Newton--Euler equation.}} At any time instant $t^{\ast}$ consider the set ${\Lambda}_{t^{\ast}}$. Let a Cartesian frame ${\mathcal E}_p$ be attached to the set under consideration. It is plain that the frame takes the same position in all sets ${\Lambda}_t$. In the frame these sets are immobile, coincide one with another and form the set noted as ${\Lambda}_p$ in the frame ${\mathcal E}_p$. We shall say that the frame ${\mathcal E}_p$ is attached to the rigid body $\alpha_p$. Let $v_{0,p}$ and $\omega _{0,p}$ be the {translation velocity} and the angular velocity of ${\mathcal E}_p$ w.r.t. ${\mathcal E}_0$.

It is plain that the vectors $\omega _{0,p}$ and $v_{0,p}$ generate the inhomogeneous slider \vspace{-3pt}
\[
	V_{0,p}=
	\{{\omega_{\hspace{.0001cm}{0},{p}}},{v_{\hspace {.0001cm}0,p}} + {\omega _{\hspace {.01cm}0,p}}{\times r_{\hspace {.0001cm}p,x}}, \forall x\in {\Lambda}_p
	\}\vspace{-3pt}
\] known as {\em kinematic}. The corresponding twist defines the reduction $V_{0,p}^{tw,p}={\rm col}\{v_{0,p}^p,\omega_{0,p}^p\}$ (here we use the fact that the vectors $\omega _{0,p}$ and $v_{0,p}$ can be considered as bounded at the point $O_p$). This reduction is called {\em vector of quasi--velocities} while its component $\omega _{0,p}^p$ is known as {\em angular quasi--velocity} \cite{Konoplev1999, Lurie}.\begin{Remark}Hereinafter one shall mark coordinate representations in any coordinate frame, {\em e.g.}, ${\mathcal E}_p$ with the help of the superscript $ {} ^ {p} $.\end{Remark}
\begin{Lemma}
There is the following relation {\rm \cite{Konoplev1999}} \vspace{3pt}
\[
{l}^{v_x,wr,p}={\it \Theta }_{p,x}^pV_{0,p}^{tw,p},\quad {\it \Theta }_{p,x}^p= 
\begin{bmatrix}
I & -r_{p,x}^{\times p} \vspace{-2pt}\\ 
 r_{p,x}^{\times p} & -(r_{p,x}^{\times p})^2
\end{bmatrix} 
\]\end{Lemma}
\proof 
The relation is true as \vspace{3pt}
\[
{l}^{v_x,wr,p} =
\begin{bmatrix}
I \vspace{-2pt}\\ 
 r_{p,x}^{\times p}
\end{bmatrix} v_x^p=
\begin{bmatrix}
I \vspace{-2pt}\\ 
 r_{p,x}^{\times p}
\end{bmatrix} (v_{0,p}^p - r_{p,x}^{\times p} \omega_{0,p}^p) = %\\ &&
\begin{bmatrix}
I & -{r}_{p,x}^{\times p} \vspace{-2pt}\\ 
 {r}_{p,x}^{\times p} & -({r}_{p,x}^{\times p} )^2
\end{bmatrix} 
\begin{pmatrix}
v_{0,p}^p \\ 
\omega_{0,p}^p
\end{pmatrix}\vspace{5pt}
\]
According to the rigid body definition the internal constraints are considered as ideal and thus \cite{Vilke} \vspace{-1pt}
	\[{\mathcal R}_{int}({\Lambda}_{t})=0 \vspace{3pt}
\]
From relations (\ref{3}) and (\ref{7}) follows that (see Appendix 2) 
\[
\int \hspace{-.1cm} \chi _{\hspace {-.05cm}_{{\Lambda_{t}}}}(
\rho_x
\frac{d}{dt} +\nu _x) l ^{\hspace{.02cm}{v}_x,wr,0}\mu_3(dx)
=\int \hspace{-.1cm} \chi _{\hspace {-.05cm}_{{\Lambda}_{t}}}
	l^{\hskip .02cm {f}_x+\xi_x,wr,0}\mu_3 (dx)\vspace{-7pt}
\]
or \vspace{-3pt}
\[\int \hspace{-.1cm} \chi _{\hspace {-.05cm}_{{\Lambda}_p}}
 {L}^{wr}_{0,p}
	[\rho _x({\it \Theta }_{p,x}^p\frac{d}{dt} +\frac{d}{dt}{\it \Theta }_{p,x}^p +\varPhi^{wr}_{0,p}{\it \Theta }_{p,x}^p) +\nu _x 
	{\it \Theta }_{p,x}^p] V_{0,p}^{tw,p}\mu_3(dx)=
	\int \hspace{-.1cm} \chi _{\hspace {-.05cm}_{{\Lambda}_p}}
	l^{\hskip .02cm {f}_x+\xi_x,wr,0}\mu_3(dx) \vspace{3pt}
 \]
where the matrices ${L}^{wr}_{0,p}$ and ${\it \varPhi}^{wr}_{0,p}$ are defined in Appendix 3.

As the twist reduction $V_{0,p}^{tw,p}$ and the matrices ${L}^{wr}_{0,p}
$ and ${\it \varPhi}^{wr}_{0,p}$ do not depend on points $x\in {\Lambda}_p$ and the matrix ${\it \Theta }_{p,x}^p $ is time--invariant, the following statement is true.
\begin{Theorem} 
The motion of $\alpha_p$ (w.r.t. ${\mathcal E}_0$ in the frame ${\mathcal E}_p$) is
described by the ({\em Newton--Euler}) equation in quasi--velocities {\rm \cite{Konoplev1999} }  
\begin{equation}{\it \Theta }^p_\rho\frac{d}{dt}\hskip .05cm V_{0,p}^{{tw,p} } +(Q_\nu^p+{\it \varPhi}^{wr}_{0,p}{\it \Theta }_\rho^p)V_{0,p}^{tw,p} ={\mathcal F}^{wr,p} 
 \label{9} \vspace{-7pt}
\end{equation}
where \vspace{-1pt}
\[\Theta^p_\rho=\hspace{-0.1cm} \int \hspace{-.1cm} \chi _{\hspace {-.05cm}_{{\Lambda}_p}}\hspace {-.1cm} {\it \Theta }_{p,x}^p \rho _x \mu_3(dx),\quad  Q^p_\nu=\hspace{-0.1cm} \int \hspace{-.1cm} \chi _{\hspace {-.05cm}_{{\Lambda}_p}} \hspace {-.1cm}{\it \Theta }_{p,x}^p \nu_x \mu_3(dx),\quad {\mathcal F}^{wr,p}=\int \hspace{-.1cm} \chi _{\hspace {-.05cm}_{{\Lambda}_p}}
	l^{\hskip .02cm {f}_x+\xi_x,wr,p}\mu_3(dx)\]
\end{Theorem}
It is easy to see that the matrices of relation (\ref{9}) depend on the rotation matrix (and translation and angular quasi--velocities, too). That is why equation (\ref{9}) must be considered along with the {Poisson kinematic relation} -- see below (\ref{21}). 

We may reduce the order of this system by using various parametric representations of rotation matrices \cite{Konoplev1999}.
\begin{Remark}
	Systems of consecutively connected rigid bodies are considered in \cite{Konoplev}.
\end{Remark}
%\vspace{-9pt} 
\subsection{A continuum} \vspace{-3pt}
{\bf \emph{Cauchy continuum}}. We shall assume that all points of ${\Lambda_t}\subset \tilde{\Lambda}_t$ are continuous. 
%Let $\mu_2$ be the restriction of $\mu_3$ on the surface ${\partial{\Lambda_t}}$, $ {n_x}$ be the normal to this surface. 

Introduce so called (homogeneous) measure ${\mathcal F}$ of mass action having the values\vspace{-5pt}
\[
	{\mathcal F}({\Lambda}_{t})=
	\int \hspace{-.1cm} \chi _{\hspace {-.05cm}_{{\Lambda_{t}}}}\rho_x l^{\hskip .02cm g_x, wr}\mu_3(dx), \quad \Lambda_{t} \subset \tilde{\Lambda}_t,\ t\in {\mathbf T} \vspace{-7pt}
\]
on the sets ${\Lambda}_{t}$ with the density ${g_x}$.

Due to \cite{Kilchevsky} constraints being in a small vicinity of $ x\in {\Lambda_t}$ cause the measure ${\mathcal R}_{int} $ of contact action or {\em stress} having the values 
\cite{Pobedria}\vspace{-3pt}
\[{\mathcal R}_{int}({\Lambda}_{t})=
	\int \hspace{-.1cm} \chi _{\hspace {-.05cm}_{\partial{\Lambda_{t}}}}l^{\hskip .02cm{P}_x, wr}%
	 n_x \mu_2(dx), \quad \Lambda_{t} \subset \tilde{\Lambda}_t,\ t\in {\mathbf T} \vspace{-3pt}\]
on the sets ${\Lambda}_{t}$ (here ${P}_x$ is {\em stress tensor}). 

 The mechanical system 
$\alpha = \{{\Lambda_t}\subset \tilde{\Lambda}_t ,\mu_3, \nu _x,\rho_x ,{\mathcal F}, {\mathcal R}_{ext}, {\mathcal R}_{int}, {\Delta\mathcal P},\forall t \in {\mathbf T}\}$, satisfying to relation (\ref{3}) with homogeneous screw measures ${\mathcal P}({\Lambda}_t)$, ${\mathcal F}$, ${\mathcal R}_{ext}$, ${\mathcal R}_{int}$ and ${\Delta\mathcal P}$, is called {\em Cauchy continuous medium} or {\em continuum}.

For the sake of brevity assume that ${\Delta\mathcal P} $ and ${\mathcal R}_{ext}\equiv 0$. 
Then due to the lemma \ref{Lemma 1} from relation (\ref{3}) follows (see Appendix 2) \vspace{-7pt}
\[(\rho _x
\frac{d}{dt} +\nu _x ){v_x}=\rho_x {g_x} %+ \xi _x     {r}_{y,x}
 +{\rm div}\hskip .05cm {P}_x, \quad {P}_x^T={P}_x\vspace{-3pt}
\]
and thus the stress tensor ${P}_x$ has to be symmetric. \medskip  
 
{\bf \emph{Continuum of Hooke class}}. 	As we will further use the divergence of ${S}_x$ and ${S}_x^{\hspace{.02cm }{}^\centerdot}$, we do not take into account their skew parts in its calculation. Denote the symmetric tensors $\frac12({S}_x+{S}_x^T)$ or $\frac12({S}_x^{\hspace{.02cm }{}^\centerdot}+{S}_x^{\hspace{.02cm }{}^\centerdot\hspace{.02cm }T})$ as ${Z}_x$.\vspace{3pt}
\begin{Definition} The Cauchy continuum is called {\em continuum of Hooke class} if the tensor ${{P}_x}$ is a linear {\em isotropic} map of ${Z}_x$, {\em i.e.}, invariant w.r.t. {Galilean group} of transformations (see also Appendix 4).%\vspace{-9pt}
\end{Definition}
%\vspace{-7pt}
\section*{Appendix 1: Sliders and screws} 
In mechanics there is mainly absent the understanding that motion of bodies and interaction between them can be described with the help of screws. It is considered as conventional that `{\em being very attractive representation of a system of forces and rigid body motions with the help motors and screws, nevertheless it has no essential practical value}' \cite{Sommerfeld} and that the screw calculus is not adapted for the description of continuum motion \cite{Dimentberg}.

Contrary to this view, we have demonstrated above that screw calculus is rather useful and convenient tools in mechanics (see also \cite{Konoplev1999} and author's paper `On Foundations of Newtonian Mechanics', arXiv:1012.3633). %However this form %fails to satisfy modern standards of explicitness as it leaves in a shade many important features of rational mechanics that can be understood only with using the (stronger) local (primitive) integral form of the conservation (change) law for the vector (screw) measure of motion (the differential form is applicable only in the cases where the divergence theorem is true). 
 %In order to obtain this form, the new notions of sliders and screw measures are used, as well as the main mechanics measures, the equation of motion and the concept of mechanical system are introduced below. 
\medskip 

{\bf \emph{Sliders}}. 
Due to the Great Soviet Encyclopedia, v. 5 (Moscow: Soviet Encyclopedia, 1971), `{\em screw calculus is the section of vector calculus in which operations over screws are studied. Here the screw %\begin{Remark}We do not support the idea to use the name `torser' from the French `torseur' instead of `screw' (Becaptures some of the features:rthelot 2006).}
 is called the pair of vectors $\{p, q\}$, which is bounded at a point $O$ and satisfied to conditions: at transition to a new point $O{\hskip .02cm}'$ the vector $ p$ does not change, and the vector $ q$ is replaced with a vector $ {q}{\hskip .075cm}' = q-\overrightarrow {(O,O{\hskip .02cm}')}\times p$ where $\times$ means cross--product. The notion of the screw is used in the mechanics (the resultant $ f$ of a force system and its main moment $ m$ form the screw $\{f, m\}$), and also in geometry (in the theory of ruled surfaces)}' (see also \cite{Berthelot}).

The definition given above is not entirely satisfactory (the screw is not the pair of vectors $\{p, q\}$), but it allows to focus on the following elements: three vectors $ p $, $ q $ and $ r $, as well as a bi--linear antisymmetric (skew) map $R(r, p) = r {\times}p$ and the relation defining the vector $ q \hspace {.05 cm} ' $ at the point $ O' $. This observation permits us to avoid the geometric constructions, which are the starting point for the conventional screw theory (see, {\em e.g.}, \cite{Dimentberg}), and to offer a simple (algebraic) version of the theory of screws. 
%On its basis it permits us, using minimal means, to consider an extensive range of issues related to the motion of material points, rigid bodies and their systems, as well as classical and non--classical (polar) continua (see also \cite{Konoplev1999}). 

Let ${\bf P}$ and ${\bf Q}$ be some (polar and pseudo, respectively) tensor fields on ${\mathbf R}^{n}$.  
A given point $x\in {\mathbf R}^{n}$ let us define the translation vector $r_{y,x}\in {\bf V}^{n}$ from a point $y\in {\mathbf R}^{n}$ to $x$. Introduce a bi--linear map $R({\bf V}^{n},{\bf P}): {\bf V}^{n}\times {\bf P}\rightarrow {\bf Q}$ as well as the following relations %\vspace{-3pt}
\begin{equation}
	{p_y }={p_x}\in {\bf P}, \quad q_y= {q_x }-	R(r_{y,x},{p_x})\in {\bf Q},\quad y \in {\mathbf R}^{n}\label{16} \vspace{11pt}
\end{equation}
\begin{Definition} The element $l ^{\hskip .02cm p_x ,q_x}=\{{p_y}\in {\bf P}, {q_y}\, {\text and}\, R(r_{y,x},{p_x}) \in {\bf Q},\forall y\in {\mathbf R}^{n}\}$ is called {\em sliding tensor--function} of $x\in {\mathbf R}^{n}$ or, briefly, {\em slider}. 
The fields ${\bf P}$ and {\bf Q} are usually called {\em resultant} and {\em moment or torque} ones, respectively \cite{Berthelot}.
\vspace{-3pt} \end{Definition}
A slider is called {\em homogeneous} if $ {q_x} =0$. In this case we shall use the notation $l^{\hskip .02cm p_x}$.%\newpage 

Denote some point of ${\mathbf R}^{n}$ by $z$. Then from (\ref{16}) follows that   \vspace{-1pt}
\[{p_z }={p_x}\in {\bf P}, \quad q_z= {q_x }-	R(r_{x,z},{p_z})\in {\bf Q} \vspace{-9pt}
\]
and 
\[{p_y }={p_z}\in {\bf P}, \quad q_y= {q_z }-	R(r_{y,z},{p_z})\in {\bf Q},\quad \forall y \in {\mathbf R}^{n}\vspace{3pt}
\]
It means that $p_z $ and $q_z $ given above can be used in order to restore sliders. \vspace{5pt}%The set of elements $l ^{\hskip .02cm p_x ,q_x}$ is an example of the field $\mathbf H$.

For the purposes of computing we may introduce slider modifications $l ^{\hskip .02cm p_x ,q_x, wr}=\{\begin{bmatrix}
{p }_y \vspace{-3pt}\\ 
q_y
\end{bmatrix}\hskip -.052cm,{p_y}\in {\bf P}, {q_y}$ and $R(r_{y,x},{p_x}) \in {\bf Q},\forall y\in {\mathbf R}^n\}$ and $l ^{\hskip .02cm p_x ,q_x, tw}\hskip -.02cm =\hskip -.02cm \{\begin{bmatrix}
{q }_y \vspace{-3pt}\\ 
p_y 
\end{bmatrix}\hskip -.052cm, {p_y}\in {\bf P}, {q_y} \ {\text and}\ R(r_{y,x},{p_x}) \in {\bf Q},\vspace{3pt} $ %\vskip .03in 

$\forall y\in {\mathbf R}^n\}$ which are called {\em wrench} and {\em twist}, respectively. For any fixed $y\in {\mathbf R}^n$, the vectors \vskip .05in 

$\begin{bmatrix}
{p }_y \vspace{-3pt}\\ 
q_y
\end{bmatrix}$ and $\begin{bmatrix}
{q }_y \vspace{-3pt}\\ 
p_y
\end{bmatrix}$ are called their {\em reductions} at the {\em reduction point} $y$. \vspace{9pt}
\medskip  
 
{\bf \emph {Screw measures (screws)}}. Below we shall use the notion of {\em signed field measure} being absolutely continuous w.r.t. $\mu _{n}$ (see also Radon--Nikodym theorem).\vspace{3pt}
\begin{Definition}Let $\sigma_{n}$ be $\sigma$--algebra on the set ${\mathbf R}^{n}$ and $\mu_{n} (A)$ be Borel measure of (\ref{1})--type. Then the following signed field measure \vspace{-5pt}
	\begin{equation}
		\Pi(A)=\int \hspace{-.1cm} \chi _{\hspace {-.05cm}_A}l^{p_x,q_x}\mu_{n} (dx),\quad {p_x}\in {\bf P}, \quad q_x \in {\bf Q}\label{17} \vspace{-3pt} 
\end{equation}
	is called {\em screw one} or, briefly, {\em screw}. 
\end{Definition}
The name {\em screw} will be used for surface integrals of sliders, too.

We assume that all discussed below sliders are $\mu _{n}$--integrable.	%Below we shall use the notion of {\em field signed measure} being absolutely continuous w.r.t. $\mu _{n}$ (see also Radon--Nikodym theorem). 
\begin{Remark}
	From relation (\ref{17}) we have \vspace{-3pt}  	
   \begin{equation}
		\int \hspace{-.1cm} \chi _{\hspace {-.05cm}_{A}}q_y\mu_{n} (dx)=\int \hspace{-.1cm} \chi _{\hspace {-.05cm}_{A}}q_x\mu_{n} (dx)-\int \hspace{-.1cm} \chi _{\hspace {-.05cm}_{A}}R(r_{y,x},f_x)\mu_{n} (dx),\quad y \in {\mathbf R}^{n}\vspace{-3pt}\label{18}
	\end{equation}
 and \vspace{-3pt}	
 \begin{equation}
		\int \hspace{-.1cm} \chi _{\hspace {-.05cm}_{A}}q_y\mu_{n} (dx)=\int \hspace{-.1cm} \chi _{\hspace {-.05cm}_{A}}q_z\mu _{n}(dx)-R(r_{y,z},\int \hspace{-.1cm} \chi _{\hspace {-.05cm}_A}f_x\mu_{n} (dx)),\quad z \in {\mathbf R}^{n}
  \label{19}\vspace{5pt}	
	\end{equation}
where the tensor $\int \hspace{-.1cm} \chi _{\hspace {-.05cm}_{A}}q_x\mu_{n} (dx)\in {\bf Q}$ would have to be called {\em intrinsic torque} of screw (\ref{17}) while the tensor $\int \hspace{-.1cm} \chi _{\hspace {-.05cm}_{A}}R(r_{y,x},f_x)\mu_{n} (dx)\in {\bf Q}$ is called {\em resultant torque}. 

We have no idea how the intrinsic torque can be defined if we might not know the tensor $q_x$ at all points of $A$. Unlike the well--known property (\ref{19}) -- see also \cite{Zhilin,Berthelot}, it is relation (\ref{18}) that is a part of the screw definition.	
	\end{Remark}
In the case where the fields ${\bf P}$ and ${\bf Q}$ may be considered as finite--dimensional vector spaces, the corresponding screws form a vector space, too %, as any screw is defined in the unique way by its reduction at some point 
 -- see also \cite{Berthelot}. 
	
In addition, we assume that all screws used below have time--independent points of reduction.\vspace{3pt}	%\medskip 
 
{\bf \emph {Sliders defined by alternants}}. 
 %\vspace{-9pt} %\medskip 
  %{\bf \emph {Newtonian mechanics as an invariant of the generalized Galilean group}}.{\bf\emph{The group ${\mathcal L}({\mathbf H}, 6)$}}. \vspace{-9pt}
 %\vspace{-7pt} \vspace{-17pt} 
Concretize the slider notion in the case where $n = 4$, ${\mathbf P}={\mathbf V}^4$, ${\mathbf Q}$ is 2nd--order
 skew $4\times 4-$pseudotensor field and $R: {\mathbf V}^4\times {\mathbf P}\rightarrow {\mathbf Q}$ is {\em alternant}: $R(r_{y,x},{p_x})=p_x\otimes r_{y,x}-r_{y,x}\otimes p_x$ where $\otimes$ means the tensor product \cite{Fedorov}.\vspace{3pt} 
 \begin{Definition} Vectors $\omega$ and $\varpi$ are called {\em dual} to a given skew 2nd--order $4\times 4-$tensor $\Omega$ if there is the following representation%\vspace{3pt}
  \[\Omega=\begin{bmatrix}
I_{3 \times 3} &\hspace{5pt} O_3 \\ 
O_3 &\hspace{5pt} O_1
\end{bmatrix} \otimes \omega +\varpi \otimes e_4-e_4\otimes \varpi%\vspace{3pt}
\]
where $I_{3 \times 3}$ is the unit tensor; $O_1$ and $O_3$ are null tensors (with corresponding orders), the vector $\varpi$ is orthogonal with the vector $e_4$ from the canonical basis ${\mathbf e}_{0}=\{e_1,e_2,e_3,e_4\}$.% of the frame ${\mathcal E}_0$. \vspace{-3pt}
 \end{Definition}
Let us show that these two vectors exist. 
Indeed, let introduce the following representations \vspace{-11pt}
\[\omega ^0 ={\rm col}\{
\omega _1 , 
\omega _2 ,
\omega_3 , 
0\}, \ \varpi ^0=
{\rm col}\{
\varpi _1 , 
\varpi _2 , 
\varpi_3 , 
0
\}, \ e_4^0=
{\rm col}\{0, 
0, 
0, 
1
\}\vspace{-5pt} 
\] 
in the basis ${\mathbf e}_{0}$. 

Then the tensor $\ \Omega\ $ has the following representation (in the basis ${\mathbf e}_{0}$) \cite{Fedorov}\vspace{-1pt}
\[\Omega^0 =\begin{bmatrix}
0 & -\omega_3 & \omega _2 & \varpi _1 \vspace{-5pt}\\ 
\omega_3 & 0 & -\omega _1 & \varpi _2 \vspace{-5pt}\\ 
-\omega _2 & \omega _1 & 0 & \varpi_3 \vspace{-5pt}\\ 
-\varpi _1 & -\varpi _2 & -\varpi_3 & 0
\end{bmatrix}\]
 Thus there exist the {\em dual} vectors $\omega$ and $\varpi$, the latter being 
orthogonal with the vector $e_4$ (note that in an other bases, one cannot guarantee that the coordinates $\omega _4$ and $\varpi _4$ are null that is why the vectors $\omega $ and $\varpi $, indeed, must be introduced as 4--dimensional). 

It is plain that \vspace{-3pt}
\[
R^0(r_{y,x},{p_x})=
\begin{bmatrix}
0 \hskip .1cm & \hskip .1cm r_2p_1-r_1p_2 \hskip .1cm & \hskip .1cmr_3p_1 -r_1p_3\hskip .1cm & \hskip .1cm-r_1p_4r_4p_1 -r_1p_4\vspace{-5pt}\\ 
r_1p_2-r_2p_1\hskip .1cm & \hskip .1cm0 \hskip .1cm & \hskip .1cmr_3p_2-r_2p_3 \hskip .1cm & \hskip .1cmr_4p_2 -r_2p_4\vspace{-5pt}\\ 
r_1p_3-r_3p_1 \hskip .1cm & \hskip .1cmr_2p_3 -r_3p_2\hskip .1cm & \hskip .1cm0 \hskip .1cm & \hskip .1cmr_4p_3 -r_3p_4\vspace{-5pt}\\ 
r_1p_4 -r_4p_1\hskip .1cm & \hskip .1cmr_2p_4 -r_4p_2\hskip .1cm & \hskip .1cmr_3p_4 -r_4p_3\hskip .1cm & \hskip .1cm0
\end{bmatrix}\vspace{-1pt}
 \]
where $p_i$ and $r_i$ are the coordinates of $p_x$ and $r_{y,x}\in {\mathbf V}^4$ ($i=\overline{1,4}$), respectively.

Hereinafter we shall denote any skew 2nd--order tensor with superscript $\ ^\times $, {\em e.g.,} $q^{\times}$.

Represent the tensors $q_y^{\times}$ and $q_x^{\times}\in {\mathbf Q}$ with the help of the matrices  \vspace{-1pt}
\[q_y^{\times 0}=\begin{bmatrix}
0 & -\alpha_3 & \alpha _2 &\overline{\alpha}_1 \vspace{-5pt}\\ 
\alpha_3 & 0 & -\alpha _1 &\overline{\alpha}_2 \vspace{-5pt}\\ 
-\alpha _2 & \alpha _1 & 0 &\overline{\alpha}_3 \vspace{-5pt}\\ 
-\overline{\alpha}_1 & -\overline{\alpha}_2 & -\overline{\alpha}_3 & 0
\end{bmatrix}, \quad q_x^{\times 0}=\begin{bmatrix}
0 & -\beta_3 & \beta _2 &\overline{\beta}_1 \vspace{-5pt}\\ 
\beta_3 & 0 & -\beta _1 &\overline{\beta}_2 \vspace{-5pt}\\ 
-\beta _2 & \beta _1 & 0 &\overline{\beta}_3 \vspace{-5pt}\\  
-\overline{\beta}_1 & -\overline{\beta}_2 & -\overline{\beta}_3 & 0
\end{bmatrix}\vspace{-1pt}\] 
Let us define the vector 
$q_y^0={\rm col}\{{\alpha}_1,{\alpha}_2,{\alpha}_3,0,\overline{\alpha}_1, \overline{\alpha}_2,\overline{\alpha}_3,0\}$ and in the same way the vector  $q_x^0$ for the tensor $q_x^{\times}$. Introduce the following relations \vspace{-3pt}
\[ 
	\begin{pmatrix}
	\hskip .1cm r_2p_1-r_1p_2 \hskip .1cm \vspace{-5pt}\\ 
	r_3p_1 -r_1p_3\vspace{-5pt}\\ 
	r_3p_2-r_2p_3 \vspace{-5pt}\\ 
	0\vspace{-7pt}\\
	r_4p_1-r_1p_4 \vspace{-5pt}\\ 
	r_4p_2 -r_2p_4\vspace{-5pt}\\ 
	r_4p_3-r_3p_4\vspace{-5pt}\\
	0
	\end{pmatrix}=\hskip .2cm
R_{y,x} ^0
	\begin{pmatrix}
	\hskip .1cm p_1 \hskip .1cm \vspace{-5pt}\\ 
	p_2 \vspace{-5pt}\\  
	p_3 \vspace{-5pt}\\ 
	p_4
	\end{pmatrix}, \quad R_{y,x} ^0 =
	\begin{bmatrix}
	0 &\hskip .1cm -r_3 &\hskip .1cm r_2 &\hskip .1cm 0 \vspace{-5pt}\\ 
r_3 &\hskip .1cm 0 &\hskip .1cm -r_1 &\hskip .1cm 0 \vspace{-5pt}\\ 
-r_2 &\hskip .1cm r_1 &\hskip .1cm 0 &\hskip .1cm 0 \vspace{-5pt}\\
	0 &\hskip .1cm 0 &\hskip .1cm 0 &\hskip .1cm 0 \vspace{-5pt}\\ 
r_4 &\hskip .1cm 0 &\hskip .1cm 0 &\hskip .1cm -r_1 \vspace{-5pt}\\ 
0 &\hskip .1cm r_4 &\hskip .1cm 0 &\hskip .1cm -r_2 \vspace{-5pt}\\ 
0 &\hskip .1cm 0 &\hskip .1cm r_4 &\hskip .1cm -r_3 \vspace{-5pt}\\
0 &\hskip .1cm 0 &\hskip .1cm 0 &\hskip .1cm 0
\end{bmatrix}%\vspace{3pt}
\]
Then due to (\ref{16}) 
 		   \[     p_y=p_x\in {\mathbf V}^4, \quad
        q_y=q_x-R_{y,x}p_x \in {\mathbf V}^8\vspace{3pt}
\]
For $n=3$ any skew 2nd--order $3\times 3-$tensor $\ \Omega\ $ is defined as 
$ \omega ^ \times=I_ {3 \times 3} \otimes \omega$ where the vector $ \omega$ is {\em dual} to $\Omega $. In the canonical basis ${\mathbf e}_{0}$ we have $\Omega^0 =\omega^{\times 0 }$ where %%\vspace{-3pt}
\[
 \omega^{\times 0 } =\begin{bmatrix}
 0 & -\omega_3 & \omega _2 \vspace{-5pt}\\ 
 \omega_3 & 0 & -\omega _1 \vspace{-5pt}\\ 
 -\omega _2 & \omega _1 & 0 
 \end{bmatrix}, \quad 
 \omega ^0 =
 \begin{pmatrix}
 \omega _1 \vspace{-5pt}\\ 
 \omega _2 \vspace{-5pt}\\ 
 \omega_3 \end{pmatrix} 
\]
Supposing that the tensor $ R(r_{y,x},{p_x})$ is an alternant
 we have %\vspace{-3pt}
 \[ R^0(r_{y,x},{p_x})=
\begin{bmatrix}
0 \hskip .1cm & \hskip .1cm r_2p_1-r_1p_2 \hskip .1cm & \hskip .1cmr_3p_1 -r_1p_3 \vspace{-5pt}\\ 
r_1p_2-r_2p_1\hskip .1cm & \hskip .1cm0 \hskip .1cm & \hskip .1cmr_3p_2-r_2p_3 \vspace{-5pt}\\ 
r_1p_3-r_3p_1 \hskip .1cm & \hskip .1cmr_2p_3 -r_3p_2\hskip .1cm & \hskip .1cm0 
\end{bmatrix}\vspace{3pt}\]
 \[  \begin{pmatrix}
	\hskip .1cm r_2p_1-r_1p_2 \hskip .1cm \vspace{-5pt}\\ 
	r_3p_1 -r_1p_3 \vspace{-5pt}\\ 
	r_3p_2-r_2p_3 
	\end{pmatrix}= 
 R_{y,x} ^0
 \begin{pmatrix}
 p_1 \vspace{-5pt}\\ 
 p_2 \vspace{-5pt}\\ 
 p_3
 \end{pmatrix}, \quad R_{y,x} ^0 =r_{y,x}^{\times 0 }=
  \begin{bmatrix}
  0 &\hskip .1cm -r_3 &\hskip .1cm r_2 \vspace{-5pt}\\ 
 r_3 &\hskip .1cm 0 &\hskip .1cm -r_1 \vspace{-5pt}\\ 
 -r_2 &\hskip .1cm r_1 &\hskip .1cm 0 
\end{bmatrix}\vspace{3pt}\]
and%\vspace{9pt}
\[ p_y=p_x\in {\mathbf V}^3, \quad
        q_y=q_x-R_{y,x}p_x=q_x-r_{y,x}\times p_x \in {\mathbf V}^3\vspace{11pt}
\]

\begin{Remark}Here we use the fact that the product $r_{y,x}^{\times 0 }\hskip .1cm p_x^{0 } $ is the coordinate representation
of the vector product $ r_{y,x} \times p_x $. %\vspace{-3pt}
\end{Remark}
From the above follows also that for $n=2$ we have  
\[	 R^0(r_{y,x},{p_x})=
\begin{bmatrix}
0 \hskip .1cm & \hskip .1cm {r_2}p_1 -{r_1}p_2\hskip .1cm \vspace{-5pt}\\ 
{r_1}p_2 -{r_2}p_1\hskip .1cm & \hskip .1cm0 \hskip .1cm 
\end{bmatrix} \vspace{-3pt}
\]
and \[
     p_y=p_x\in {\mathbf V}^2, \quad
        q_y=q_x-R_{y,x}p_x \in {\mathbf V}^1
        \] where $  R_{y,x} ^0 =\begin{bmatrix}
-r_2, & r_1
\end{bmatrix} 
$.\vspace{7pt}
\begin{Remark}In the above we could use the 2nd--order tensor field $\ \bf P$ instead of ${\mathbf V}^4$.
\end{Remark}\vspace{-9pt}
\section*{Appendix 2: Derivatives of some integrals defined over {actual shapes} $\Lambda_{t}$ of mechanical systems} 
We shall assume that Stocks theorem is applicable and there are the following statements (see also \cite{Pobedria}).
\begin{Lemma}
	\label{26}Let $f (x, t)$ be a measurable function on some set $G\in {\mathbf R}^3$. If \vspace{-5pt}
	\[\int \hspace{-.1cm} \chi _{\hspace {-.05cm}_{V}} \hspace {-.05cm}f (x, t)\mu _3(dx)=0\vspace{-3pt}\]
	for any subset $V\in G$ and any $t\in {\mathbf T}$, then $f (x, t)\equiv 0$ in $G$.\vspace{7pt}
\end{Lemma}
\begin{Lemma}\label{Lemma 02}For any measurable function $f (x, t)$ on $\Lambda_{t} \subset \tilde{\Lambda}_t$, $t\in {\mathbf T}$, we have\vspace{-3pt} \[
	\frac{d}{dt}\hspace{-.05cm} \int \hspace{-.1cm} \chi _{\hspace {-.05cm}_{\Lambda_{t}}} f (x, t) \mu _3(dx) =
				\int \hspace{-.1cm} \chi _{\hspace {-.05cm}_{{\Lambda}_{t}^{ac}}}\hspace {.05cm}[\frac{d}{dt}f (x, t) + f (x, t) {\rm div}\hskip.05cm {v_x}]
\hskip.05cm \mu_{ac}(dx)\hspace {-.035cm}+ \sum_k \frac{d}{dt}\hspace {.05cm} f ({x_k},t)\hskip.05cm \mu_{pp}(x_k)\vspace{5pt}
\]\end{Lemma}
 \begin{Lemma}\label{Lemma 03}For any measurable function $f (x, t)$ on $\Lambda_{t} \subset \tilde{\Lambda}_t$, $t\in {\mathbf T}$, we have %\vspace{1pt}
\[
	\frac{d}{dt}\hspace{-.05cm} \int \hspace{-.1cm} \chi _{\hspace {-.05cm}_{\Lambda_{t}}} \rho_x f (x, t) \mu _3(dx) =
				\int \hspace{-.1cm} \chi _{\hspace {-.05cm}_{{\Lambda}_{t}^{ac}}}\hspace {.05cm}(\rho_x \frac{d}{dt} +\nu _{x})
\hskip.1cm f (x, t) \mu_{ac}(dx)\hspace {-.035cm}+ \sum_k (\rho_{k}\frac{d}{dt}+\nu _{k} )\hspace {.1cm} f ({x_k},t)\hskip.05cm \mu_{pp}(x_k)\vspace{5pt}
\]
\end{Lemma}
We assume that all integrals above have sense in the above relations.
\vspace{-9pt}\section*{Appendix 3: The generalized Galilean group}\vspace{-9pt}
Consider %a Galilean frame of reference ${\it \varphi}_0$ and 
 two Cartesian frames ${\mathcal E}_0$ and ${\mathcal E}_p$ in ${\mathbf R}^3$ with bases 
 ${\mathbf e}_0$ and ${\mathbf e}_p$, respectively. We shall assume that the former is immobile while the latter can move w.r.t. the former. Introduce the radius--vectors ${r}_x$ and ${r}_{p,x}$ of a point $x\in {\mathbf R}^3$ w.r.t. the origins $O_0$ and $O_p$, respectively. Define the vector $d_{0,p}=r_x-{r_{p,x}}$. The vector 
${v_x}=r_x^{{}^\centerdot }$ is the velocity of $x$ w.r.t. $O_0$ while the vector ${v_{0,p}}=d_{0,p}^{{\hspace{.04cm }}^\centerdot }$ is {\em translation velocity} of ${\mathcal E}_p$ w.r.t. $O_0$. 
 
There exists a unique vector $\omega_{0,p}$ such that \cite{Yakovenko} 
\[{v_x}=v_{0,p}+\omega_{0,p}\times r_{p,x}, \quad \forall x\in {\mathbf R}^3 
\] 	
We may represent the relation $r_x=d_{0,p}+{r}_{p,x}$ in the coordinate frame ${\mathcal E}_0$ as $r^0_x=d^0_{\hspace{.02cm }0,p}+C_{0,p}r_{p,x}^p$. With differentiating the above relation we have 
$v^0_x=v_{0,p}^{0}+C_{0,p}^{\hspace{.02cm }{}^\centerdot}r_{p,x}^p=v_{0,p}^{0}+C_{0,p}^{\hspace{.02cm }{}^\centerdot}C_{p, 0}r_{p,x}^0$. 
 
In the coordinate frame ${\mathcal E}_p$ we have $v^p_x=v_{0,p}^{p}+C_{p, 0}C_{0,p}^{\hspace{.02cm }{}^\centerdot}r_{p,x}^p$. 
Thus the entries of the cross--product matrix \vspace{-3pt}
\begin{equation}
		\omega _{0,p}^{\times p} \stackrel{\rm {def}}{=}C_{p, 0}C_{0,p}^{\hspace{.02cm }{}^\centerdot}
				\label{20}\vspace{-7pt}
\end{equation}
 define the coordinates of $\omega _{0,p}$ in ${\mathcal E}_p$. 

From (\ref{20}) follows the {\em Poisson kinematic relation} \vspace{-3pt}
\begin{equation}
 C_{0,p}^{\hspace{.02cm }{}^\centerdot}=C_{0,p}\omega _{0,p}^{\times p}
\label{21}\vspace{-3pt}\end{equation} 
Introduce the following matrices  \vspace{3pt}
\begin{equation}
{C}_{0,p}^\oplus 
= 
\begin{bmatrix}
C_{0,p} &\ O \vspace{-5pt}\\ 
O &\ C_{0,p}
\end{bmatrix}, \quad 
{D}^0_{0,p}=
\begin{bmatrix}
I & \hspace{.05cm} O \vspace{-5pt}\\ 
 \hspace{.05cm}d_{0,p}^{\hspace{.03cm }\times 0} & \hspace{.05cm}I
\end{bmatrix}, \quad 
{D}^p_{0,p}=
\begin{bmatrix}
I &\ O \vspace{-5pt}\\ 
 \hspace{.15cm}d_{0,p}^{\hspace{.03cm }\times p} &\ \hspace{.15cm}I
\end{bmatrix}
\label{22} \vspace{3pt}
 \end{equation}
	where $I$ is the unit matrix, $O$ is the zero one.
%Here the superscripts $\ {}^{\times 0}\ $ and $\ {}^{\times p}\ $ note cross--product matrices generated by coordinates of vectors in ${\mathcal E}_0$ and ${\mathcal E}_p$, respectively.	

\begin{Theorem} Let $\mathbf H $ be the set of all sliders.
 A given inhomogeneous slider $l\in \mathbf H$\vspace{-3pt}
\[l ^{wr, 0}= {L}^{wr}_{0,p}l^{wr , p}\vspace{-3pt}\] 
where
$l ^{wr, 0}$ and $l ^{wr , p}$ are wrench reductions of the slider $l$ computed in the bases 
 ${\mathbf e}_0$ and ${\mathbf e}_p$, respectively, the matrix ${L}^{wr}_{0,p}$ is defined by the following relation {\rm \cite{Konoplev1999}} 
\begin{equation}
	{L}^{wr}_{0,p}={C}_{0,p}^\oplus {D}^p_{0,p}={D}^0_{0,p}{C}_{0,p}^\oplus \label{23}
\end{equation}
 and belongs to the multiplicative group ${\mathcal L}^{wr}$ such that 
\begin{equation} 
{L}^{{wr}\centerdot }_{0,p}={L}^{wr}_{0,p}{\it \varPhi}_{0,p}^{wr}={\it \varPsi}_{0,p}^{wr}{L}^{wr}_{0,p},
\ {\it \varPhi}_{0,p}^{wr}=
\begin{bmatrix}
 \omega_{0,p}^{\times p} &\ O \vspace{3pt}\\ 
 v_{0,p}^{\times p} &\ \omega_{0,p}^{\times p}
\end{bmatrix}, \ {\it \varPsi}_{0,p}^{wr}=
\begin{bmatrix}
 \omega_{0,p}^{\times 0} &\ O \vspace{3pt}\\ 
 v_{0,p}^{\times 0} &\ \omega_{0,p}^{\times 0}
\end{bmatrix}
\label{24}%\vspace{3pt}
\end{equation}
\end{Theorem}
\proof Relation (\ref{23}) follows directly from the slider definition.

Consider the case where ${L}^{{wr}}_{0,p}={C}_{0,p}^\oplus {D}^p_{0,p}$. Then from (\ref{22}) follows that 
 ${L}^{{wr}\centerdot}_{0,p}={C}_{0,p}^{\oplus \centerdot} {D}^p_{0,p}+{C}_{0,p}^\oplus {D}^{p\centerdot}_{0,p}=({C}_{0,p}^{\oplus \centerdot} {D}^p_{0,p}{D}^p_{p,0}{C}_{p,0}^\oplus +{C}_{0,p}^\oplus {D}^{p\centerdot}_{0,p}{D}^p_{p,0}{C}_{p,0}^\oplus ){C}_{0,p}^\oplus {D}^p_{0,p}
=({C}_{0,p}^{\oplus \centerdot}{C}_{0,p}^\oplus +{D}^{0\centerdot}_{0,p}){C}_{0,p}^\oplus {D}^p_{0,p}={\it \varPsi}_{0,p}^{wr}{L}^{wr}_{0,p}$. 

In the case where ${L}^{{wr}}_{0,p}={D}^0_{0,p}{C}_{0,p}^\oplus$, from (\ref{22}) follows that 
${L}^{{wr}\centerdot}_{0,p}={D}^{0\centerdot}_{0,p}C_{0,p}^\oplus +{D}^0_{0,p}{C}^{\oplus \centerdot}_{0,p}={D}^0_{0,p}
{C}_{0,p}^\oplus ({C}_{p,0}^\oplus {C}^{\oplus \centerdot}_{0,p}$ $+{C}_{p,0}^\oplus {D}^0_{p,0}{D}^{0\centerdot
}_{0,p}{C}_{0,p}^\oplus ) ={L}^{wr}_{0,p}{\it \varPhi}_{0,p}^{wr}$. Thus we have relation (\ref{24}).

Let ${C}_{p,k}$ be the rotation matrix of a Cartesian frame ${\mathcal E}_k$ w.r.t. ${\mathcal E}_p$. Then $L^{{wr}}_{0,p}L^{{wr}}_{p,k}={C}_{0,p}^\oplus {C}_{p,k}^\oplus {C}_{k, p}^\oplus {D}^p_{0,p}{C}_{p,k}^\oplus {D}^k_{p,k}$ $= 
{C}_{0,k}^\oplus {D}^k_{0,p}{D}^k_{p,k}={C}_{0,k}^\oplus {D}^k_{0,k}=L^{{wr}}_{0,k}$ and $%
L_{0,p}^{{{wr}}, - 1}=({C}_{0,p}^\oplus {D}^p_{0,p})^{- 1}=(D^p_{0,p})^{- 1}C_{0,p}^{\oplus, T}=D^p_{p,0}C_{p,0}^{\oplus}
=C_{p,0}^{\oplus}D^0_{p,0}=L^{{wr}}_{p,0}$, {\em i.e.}, matrices of the kind ${L}^{wr}_{0,p}={C}_{0,p}^\oplus {D}^p_{0,p}$ 
form a multiplicative group.\vspace{5pt}

The similar statement $l ^{tw, p}= {L}^{tw}_{0,p}l^{tw ,0}$ is true for twists where we have the matrix \vspace{-1pt} 
\begin{equation}
	{L}^{tw}_{0,p}=\begin{bmatrix} O & \hspace {.25cm} I  \\ I 
	&\hspace {.25cm} O \end{bmatrix}{L}^{wr}_{0,p}\begin{bmatrix} O & \hspace {.25cm} I \\ I 	&\hspace {.25cm} O \end{bmatrix}\label{25}%\vspace{5pt}
\end{equation} 
belongs to the multiplicative group ${\mathcal L}^{tw}$ such that $
{L}^{{tw}\centerdot }_{0,p}={\it \varPsi}_{0,p}^{tw}{L}^{tw}_{0,p}$, 
$ {\it \varPsi}_{0,p}^{tw}=- {\it \varPsi}_{0,p}^{wr, T}
$ and $ 
{L}^{{tw}\centerdot }_{0,p}={\it \varPhi}_{0,p}^{tw}{L}^{tw}_{0,p}$, 
$ {\it \varPhi}_{0,p}^{tw}=- {\it \varPhi}_{0,p}^{wr, T}
$. 
%It is plain that the groups ${\mathcal L}^{wr}$ and ${\mathcal L}^{tw}$ generate the same group ${\mathcal L}({\mathbf H},6)$.

The groups ${\mathcal L}^{wr}$ and ${\mathcal L}^{tw}$ generate the group ${\mathcal L}({\mathbf H}, 6)$ acting in the slider set ${\mathbf H}$ when there is a change of coordinate frames, {\em e.g.}, from ${\mathcal E}_0$ to ${\mathcal E}_p$.\vspace{5pt}
\begin{Definition}
	The group ${\mathcal {GL}}({\mathbf H}, 6)\subset {\mathcal L}({\mathbf H}, 6)$ is called {\em generalized Galilean group} if the translation vector $d_{0,p}$ is replaced with $d_{0,p}+v_{0,p} t$, the initial translation $d_{0,p}$ and the translation velocity $v_{0,p}$ of ${\mathcal E}_p$ w.r.t. ${\mathcal E}_0$ are constant while the angular velocity $\omega _{0,p}$ is constant, too \cite{Konoplev1999}. 
\end{Definition}%\vspace{-9pt}
\vspace{-9pt}
\section*{Appendix 4: Multiplicative groups of linear isotropic maps}
{\bf {\emph {3--dimensional case}}}. \vspace{-3pt}
 Given any $3\times 3$--matrix $Z$, let us define the linear independent matrices
\begin{equation}
	E_1=({\rm trace \hskip .02in}{\it Z})I,\ E_2={\it Z}, \ E_2={\it Z}^T \label{28}\vspace{3pt}
\end{equation}
Aggregates $AE_iB$ ($i=\overline{1,3}$) are isotropic functions of $Z$ entries if the matrices $A$ and $B$ are
proportional to $I$ with scalar coefficients being invariant w.r.t. {Galilean group} of transformations. Hereinafter all scalar coefficients used below will be considered as invariant w.r.t. {this group}.

Let us construct the sets of all isotropic functions \cite{Dubrovin}\vspace{-1pt}
\[
P=p_0I+p_1({\rm trace \hskip .02in}Z) I+p_2Z+p_3Z^T,
 \quad 
 Q=q_0I+q_1({\rm trace \hskip .02in}P) I+q_2P+q_3P^T\vspace{-3pt} \]
where $p_i$ and $q_i$ are scalar coefficients.

Then \vspace{-7pt}
\begin{eqnarray*}
 Q&=&q_0I+q_1[3p_0+(3p_1+p_2+p_3){\rm trace \hskip .02in}Z]I+q_2[p_0I+p_1(
{\rm trace \hskip .02in}Z) I+p_2Z+p_3Z^T]+\\
&& q_3[p_0I+p_1({\rm trace \hskip .02in}Z) I+p_2Z^T+p_3Z]\\
&=&(q_0+3p_0q_1+p_0q_2+p_0q_3)I+[(3p_1+p_2+p_3)q_1+p_1q_2+p_1q_3]{\rm 
(trace \hskip .02in}Z) I +\\
&&(p_2q_2+p_3q_3)Z+(p_3q_2+p_2q_3)Z^T\\
&=&r_0I+r_1({\rm trace \hskip .02in}Z) I+r_2Z+r_3Z^T\vspace{-9pt}
\end{eqnarray*}
where %
\[\begin{pmatrix}
r_0 \vspace{-5pt}\\ 
r_1 \vspace{-5pt}\\ 
r_2 \vspace{-5pt}\\ 
r_3
\end{pmatrix}=R
\begin{pmatrix}
q_0 \vspace{-5pt}\\ 
q_1 \vspace{-5pt}\\ 
q_2 \vspace{-5pt}\\ 
q_3
\end{pmatrix}, \quad R=\begin{bmatrix}
1 & 3p_0 & p_0 & p_0 \vspace{-5pt}\\ 
0 & 3p_1+p_2+p_3 & p_1 & p_1 \vspace{-5pt}\\ 
0 & 0 & p_2 & p_3 \vspace{-5pt}\\ 
0 & 0 & p_3 & p_2
\end{bmatrix}\vspace{3pt}\]
The set of non--singular matrices of the kind $R$ forms 
 a multiplicative group. That is why if ${\rm det }\ R=(3p_1+p_2+p_3) (p_2^2-p_3^2)\neq 0$
then in the case where $Q=Z$ we have  
\[\begin{pmatrix}
q_0 \vspace{-5pt}\\ 
q_1 \vspace{-5pt}\\ 
q_2 \vspace{-5pt}\\ 
q_3
\end{pmatrix} =R^{-1}\begin{pmatrix}
0 \vspace{-5pt}\\ 
0 \vspace{-5pt}\\ 
1 \vspace{-5pt}\\ 
0
\end{pmatrix} =\begin{pmatrix}
\frac{-p_0}{3p_1+p_2+p_3} \\ 
\frac{-p_1}{(p_2+p_3) (3p_1+p_2+p_3) } \\ 
\frac{p_2}{p_2^2-p_3^2} \\ 
\frac{-p_3}{p_2^2-p_3^2}
\end{pmatrix}\vspace{3pt}\]
and the inverse function $
Z=q_0I+q_1({\rm trace \hskip .02in}P) I+q_2P+q_3P^T 
$.
\begin{Remark}Instead 	(\ref{28}) we might choose an other set of linear independent matrices, {\em e.g.}, $E_1=({\rm trace \hskip .02in}{\it Z})I,\ E_2=\frac12({\it Z}+{\it Z}^T), \ E_2=\frac12({\it Z}-{\it Z}^T)$. 
\end{Remark}
In the case where the matrix $Z$ is symmetric let us define the linear independent matrices $E_1=({\rm trace \hskip .02in}{\it Z})I,\ E_2={\it Z}$ and the sets of linear isotropic functions
 $P=p_0\hskip .05cm{I}+p_1\hskip .05cm({\rm trace \hskip .02in}{\it Z})\hskip .05cm I+{\it r}_2{\it Z}%+{\it r}^3{\it Z}^{T}	
$, $Q=q_0I+q_1({\rm trace \hskip .02in}P) I+q_2P\vspace{3pt}$. 
As ${\rm trace \hskip .02in}P=3p_0+(3p_1+p_2) {\rm trace \hskip .02in}Z$ we have\vspace{-5pt}
\[
Q =(q_0+3p_0q_1+p_0q_2)I+[(3p_1+p_2)q_1+p_1q_2
]({\rm trace \hskip .02in}Z) I +q_2[p_0\hskip .05cm{I}+p_1\hskip .05cm({\rm trace \hskip .02in}{\it Z})\hskip .05cm I+{\it r}_2{\it Z}
]\vspace{-3pt}\]
or $Q=r_0I+r_1({\rm trace \hskip .02in}Z) I+r_2Z$ where
\[
\begin{pmatrix}
r_0 \vspace{-5pt}\\ 
r_1 \vspace{-5pt}\\ 
r_2 
\end{pmatrix}
=R\begin{pmatrix}
q_0 \vspace{-5pt}\\ 
q_1 \vspace{-5pt}\\ 
q_2 
\end{pmatrix}, \quad R=
	\begin{bmatrix}
1 & 3p_0 & p_0 \vspace{-5pt}\\ 
0 & 3p_1+p_2 & p_1 \vspace{-5pt}\\ 
0 & 0 & p_2
\end{bmatrix} 
\] 
Thus under the condition that ${\rm det }\ R=(3p_1+p_2) p_2\neq 0$ for the case where $Q=Z$ we have 
\[\begin{pmatrix}
q_0 \vspace{-5pt}\\ 
q_1 \vspace{-5pt}\\ 
q_2 
\end{pmatrix} =R^{-1}\begin{pmatrix}
0 \vspace{-5pt}\\ 
0 \vspace{-5pt}\\ 
1
\end{pmatrix} =\begin{pmatrix}
\frac{-p_0}{3p_1+p_2} \\ 
\frac{-p_1}{p_2 (3p_1+p_2) } \\ 
\frac{1}{p_2} 
\end{pmatrix}\]
and the inverse function $
Z=q_0I+q_1({\rm trace \hskip .02in}P) I+q_2P$. \vspace{3pt}
\medskip

{\bf {\emph {2--dimensional case}}}. %\vspace{7pt}
Since the formulation of the axioms of mechanics can be restated in the two--dimensional space and Lemma \ref{Lemma 1} can be reformulated with using Green's theorem, let us consider the multiplicative group of isotropic maps for 2--dimensional media.\vspace{3pt}

Given $2\times 2$--matrix $Z=\begin{bmatrix}
a\ & b \vspace{-5pt}\\ 
c\ & d
\end{bmatrix} $ let us define the matrices $E_1=({\rm trace \hskip .02in}{\it Z})I$, 
${E}_2=({\rm trace \hskip .02in}\hskip .025cm\widetilde{I}Z){I}$,\vspace{3pt}

$E_3={Z}$,
${E}_4=\widetilde{I}{Z}$, 
$E_5={Z}^T$, 
${E}_6={Z}^T\widetilde{I}$, 
${E}_7=\widetilde{I}{Z}^T$, 
${E}_8={Z}\widetilde{I}$, 
${E}_9=\widetilde{I}{Z}\widetilde{I}$ 
and 
${E}_{10}=\widetilde{I}{Z^T}\widetilde{I}$\vspace{3pt}

 where $I$ is the identity 	$2\times 2$ matrix, $\widetilde{I}=
\begin{bmatrix}
0 & -1 \vspace{-5pt}\\ 
1 & 0
\end{bmatrix}$.\vspace{5pt}

It is plain that the first 6 matrices $E_1$, ${E}_2$, $E_3$, ${E}_4$, $E_5$ and ${E}_6$ are linear independent and ${E}_7={E}_4-{E}_2$,
${E}_8={E}_2+{E}_6$, 
${E}_9=-E_1+E_5$, 
${E}_{10}=-E_1+E_3$.

Aggregates $AE_iB$ are isotropic maps of $Z$ entries if $A$ and $B$ are
of the kind $\alpha I+\beta\widetilde{I}$ where $\alpha $ and $\beta$ are scalar coefficients.

Introduce the sets of isotropic functions \vspace{-3pt}
\begin{eqnarray*}P&=&p_0I+\widetilde{p}_0\widetilde{I}+p_1({\rm trace \hskip .02in}Z) I+p_2 
({\rm trace \hskip .02in}\widetilde{I}Z)I+p_3Z+p_4Z^T+p_5\widetilde{I}Z+p_6Z^T
\widetilde{I}\\
Q&=&q_0I+\widetilde{q}_0\widetilde{I}+q_1({\rm trace \hskip .02in}P) I+q_2
({\rm trace \hskip .02in}\widetilde{I}P)I+q_3P+q_4P^T+q_5\widetilde{I}P+q_6P^T
\widetilde{I}
 \end{eqnarray*}
Then\vspace{-3pt}%\newpage
\begin{eqnarray*}
{\rm trace \hskip .02in}P&=&2p_0+(2p_1+p_3+p_4){\rm trace \hskip .02in}Z+(2p_2+p_5-p_6){\rm 
trace \hskip .02in}\widetilde{I}Z
\\
\widetilde{I}P&=&p_0\widetilde{I}-\widetilde{p}_0I+p_1( {\rm trace \hskip .02in} 
Z) \widetilde{I}+p_2({\rm trace \hskip .02in}\widetilde{I}Z)\widetilde{I}+p_3 
\widetilde{I}Z+p_4\widetilde{I} Z-p_5Z+
\\
&&p_6\widetilde{I}Z^T\widetilde{I} 
\\{\rm trace \hskip .02in}\widetilde{I}P&=&-2\widetilde{p}_0-(p_5+p_6){\rm 
trace \hskip .02in}Z +(p_3-p_4) {\rm trace \hskip .02in}\widetilde{I}Z\\
P^T\widetilde{I} &=&p_0\widetilde{I}+\widetilde{p}_0I+p_1( {\rm 
trace \hskip .02in}Z) \widetilde{I}+p_2({\rm trace \hskip .02in}\widetilde{I}Z)\widetilde{I} 
+p_3Z^T\widetilde{I}+ p_4Z\widetilde{I}+p_5Z^T-\\&&
p_6\widetilde{I}Z\widetilde{I} \vspace{-7pt}
\end{eqnarray*}
As 
 $\widetilde{I}Z^T=\widetilde{I}Z-( {\rm trace \hskip .02in}\widetilde{I}Z)I$, 
$Z\widetilde{I}=({\rm trace \hskip .02in}\widetilde{I}Z)I+Z^T\widetilde{I}$, 
$\widetilde{I}{Z}\widetilde{I}=Z^T-( {\rm trace \hskip .02in}Z) I$, 
$\widetilde{I}Z^T\widetilde{I}=Z-( {\rm trace \hskip .02in}Z) I$, 
${\rm trace \hskip .02in}Z=a+d$, ${\rm trace \hskip .02in}\widetilde{I}Z=b-c$, ${\rm trace \hskip .02in}\widetilde{I}Z^T=-(b-c)$, 
$( {\rm trace \hskip .02in}Z) \widetilde{I}=\widetilde{I}Z+Z^T\widetilde{I}$, 
${\rm trace \hskip .02in}\widetilde{I}Z=-{\rm trace \hskip .02in}Z^T\widetilde{I}$, 
${\rm trace \hskip .02in}Z\widetilde{I}={\rm trace \hskip .02in}\widetilde{I}Z=b-c$, 
$({\rm trace \hskip .02in}Z\widetilde{I})\widetilde{I}=Z^T-Z, ({\rm trace \hskip .02in}\widetilde{I}Z)\widetilde{I}=Z^T-Z=({\rm trace \hskip .02in}Z
\widetilde{I})\widetilde{I}$, 
${\rm trace \hskip .02in}\widetilde{I}Z^T\widetilde{I}={\rm trace \hskip .02in}\widetilde{I}Z 
\widetilde{I}=- {\rm trace \hskip .02in}Z$ we have %\newpage
\begin{eqnarray*}
Q&=&q_0I+\widetilde{q}_0\widetilde{I}+ 
[2p_0q_1+(2p_1+p_3+p_4)q_1{\rm trace \hskip .02in}Z+(2p_2+p_5-p_6)q_1 
({\rm trace \hskip .02in}\widetilde{I}Z)]I\\
&&+[-2\widetilde{p}_0q_2-(p_5+p_6)q_2( {\rm trace \hskip .02in}Z)
+(p_3-p_4)q_2( {\rm trace \hskip .02in}\widetilde{I}Z) ]I+ \\
&&p_0q_3I+\widetilde{p}_0q_3\widetilde{I}+p_1q_3( {\rm trace \hskip .02in} 
Z) I+p_2q_3({\rm trace \hskip .02in}\widetilde{I}Z)I+p_3q_3Z+p_4q_3Z^T+
\\&&p_5q_3\widetilde{I}Z+p_6q_3Z^T\widetilde{I}+\\
&&p_0q_4I-\widetilde{p}_0q_4\widetilde{I}+p_1q_4( {\rm trace \hskip .02in} 
Z) I+p_2q_4({\rm trace \hskip .02in}\widetilde{I}Z)I+p_4q_4Z+p_3q_4Z^T-
\\&& p_6q_4\widetilde{I}Z-p_5q_4Z^T\widetilde{I}+\\
&&p_0q_5\widetilde{I}-\widetilde{p}_0q_5I-p_6q_5( {\rm trace \hskip .02in} 
Z) I-p_4q_5({\rm trace \hskip .02in}\widetilde{I}Z)I+(p_6-p_5-p_2)q_5Z+ 
\\&&p_2q_5Z^T+(p_1+p_3+p_4)q_5\widetilde{I}Z+p_1q_5Z^T\widetilde{I} +\\
&&p_0q_6\widetilde{I}+\widetilde{p}_0q_6I+p_6q_6({\rm trace \hskip .02in} 
Z) I+p_4q_6({\rm trace \hskip .02in}\widetilde{I}Z)I-p_2q_6Z+
\\&&(p_2+p_5-p_6)q_6Z^T+p_1q_6\widetilde{I}Z+(p_1+p_3+p_4)q_6Z^T\widetilde{I}\vspace{-3pt}
\end{eqnarray*}
or \vspace{-3pt}
\begin{eqnarray*}
Q&=&(q_0+2p_0q_1-2\widetilde{p}_0q_2+p_0q_3+p_0q_4-\widetilde{p}_0q_5+ 
\widetilde{p}_0q_6)I+ 
\\
&& (\widetilde{p}_0+\widetilde{p}_0q_3-\widetilde{p} 
_0q_4+p_0q_5+p_0q_6)\widetilde{I}+ 
\\&&
[(2p_1+p_3+p_4)q_1-(p_5+p_6)q_2+p_1q_3+p_1q_4-p_6q_5+p_6q_6] {\rm trace \hskip .02in}Z+\\
&&[(2p_2+p_5-p_6)q_1+(p_3-p_4)q_2+p_2q_3+p_2q_4-p_4q_5+p_4q_6]{\rm trace \hskip .02in}\widetilde{I}Z+
\\ &&[p_3q_3+p_4q_4+(p_6-p_5-p_2)q_5-p_2q_6]Z+\\
&& 
[p_4q_3+p_3q_4+p_2q_5+(p_2+p_5-p_6)q_6]Z^T+\\
&&[p_5q_3-p_6q_4+(p_1+p_3+p_4)q_5+p_1q_6]\widetilde{I}Z+
\\&&
[p_6q_3-p_5q_4+p_1q_5+(p_1+p_3+p_4)q_6]Z^T\widetilde{I}\\
&=&
r_0I+\widetilde{r}_0\widetilde{I}+r_1( {\rm trace \hskip .02in}Z) I+r_2 
({\rm trace \hskip .02in}\widetilde{I}Z)I+r_3Z+r_4Z^T+r_5\widetilde{I}Z+r_6Z^T 
\widetilde{I} 
\end{eqnarray*}
Hence \[{\rm col}\{r_0,\widetilde{r}_0,r_1,r_2,r_3,r_4,r_4,r_5,r_6\}=R\ {\rm col}\{q_0,\widetilde{q}_0,q_1,q_2,q_3,q_4,q_4,q_5,q_6\}\vspace{-3pt}\] where \vspace{-3pt}%\newpage
\[R=
\begin{bmatrix}
1\ & 0 & 2p_0 & -2\widetilde{p}_0 & p_0 & p_0 & -\widetilde{p}_0 & \widetilde{
r}_0 \vspace{-5pt}\\ 
0\ & 1 & 0 & 0 & \widetilde{p}_0 & -\widetilde{p}_0 & p_0 & p_0 \vspace{-5pt}\\ 
0\ & 0 & 2p_1+p_3+p_4 & -(p_5+p_6) & p_1 & p_1 & -p_6 & p_6 \vspace{-5pt}\\ 
0\ & 0 & 2p_2+p_5-p_6 & p_3-p_4 & p_2 & p_2 & -p_4 & p_4 \vspace{-5pt}\\ 
0\ & 0 & 0 & 0 & p_3 & p_4 & p_6-p_5-p_2 & -p_2 \vspace{-5pt}\\ 
0\ & 0 & 0 & 0 & p_4 & p_3 & p_2 & p_2+p_5-p_6 \vspace{-5pt}\\ 
0\ & 0 & 0 & 0 & p_5 & -p_6 & p_1+p_3+p_4 & p_1 \vspace{-5pt}\\ 
0\ & 0 & 0 & 0 & p_6 & -p_5 & p_1 & p_1+p_3+p_4
\end{bmatrix} 
\]
It is plain that 
${\rm det }\ R =-4[(2p_1+p_3+p_4)(p_3-p_4)+(2p_2+p_5-p_6)(p_5+p_6)]^2[(p_3+p_4)^2+(p_5-p_6)^2] $. Under the condition that ${\rm det }\ R\neq 0$ in the case where $Q=Z$ we have \vspace{-3pt}%\newpage
\[\begin{pmatrix}
q_0 \vspace{-5pt}\\ 
\widetilde{q}_0 \vspace{-5pt}\\ 
q_1 \vspace{-5pt}\\ 
q_2
\end{pmatrix} =-\begin{bmatrix}
1\ & 0 & 2p_0 & -2\widetilde{p}_0 \vspace{-5pt}\\ 
0\ & 1 & 0 & 0 \vspace{-5pt}\\ 
0\ & 0 & 2p_1+p_3+p_4 & -(p_5+p_6) \vspace{-5pt}\\ 
0\ & 0 & 2p_2+p_5-p_6 & p_3-p_4
\end{bmatrix} ^{-1}\begin{bmatrix}
p_0 & p_0 & -\widetilde{p}_0 & \widetilde{p}_0 \vspace{-5pt}\\ 
\widetilde{p}_0 & -\widetilde{p}_0 & p_0 & p_0 \vspace{-5pt}\\ 
p_1 & p_1 & -p_6 & p_6 \vspace{-5pt}\\ 
p_2 & p_2 & -p_4 & p_4
\end{bmatrix} \begin{pmatrix}
q_3 \vspace{-5pt}\\ 
q_4 \vspace{-5pt}\\ 
q_5 \vspace{-5pt}\\ 
q_6
\end{pmatrix}\]
\[
\begin{pmatrix}
q_3 \vspace{-5pt}\\ 
q_4 \vspace{-5pt}\\ 
q_5 \vspace{-5pt}\\ 
q_6
\end{pmatrix}=\begin{bmatrix}
p_3 & p_4 & p_6-p_5-p_2 & -p_2 \vspace{-5pt}\\ 
p_4 & p_3 & p_2 & p_2+p_5-p_6 \vspace{-5pt}\\ 
p_5 & -p_6 & p_1+p_3+p_4 & p_1 \vspace{-5pt}\\ 
p_6 & -p_5 & p_1 & p_1+p_3+p_4
\end{bmatrix}^{-1}\begin{pmatrix}
1 \vspace{-5pt}\\ 
0 \vspace{-5pt}\\ 
0 \vspace{-5pt}\\ 
0
\end{pmatrix}\] and the inverse function 
\[
Z=q_0I+\widetilde{q}_0\widetilde{I}+q_1({\rm trace \hskip .02in}P) I+q_2 
({\rm trace \hskip .02in}\widetilde{I}P)I+q_3P+q_4P^T+q_5\widetilde{I}P+q_6P^T 
\widetilde{I} 
\]
If the matrix $Z$ is symmetric we have the linearly independent matrices $E_1= {\rm trace \hskip .02in}Z$, ${E}_2=Z$ and $E_3=\widetilde{I}Z-Z\widetilde{I}$. Define the sets of linear isotropic functions
\begin{eqnarray*}
P&=&p_0I+p_1( {\rm trace \hskip .02in}Z) I+p_2Z+p_3(\widetilde{I}Z-Z\widetilde{I})\\
Q&=&q_0I+q_1( {\rm trace \hskip .02in}P) I+q_2P+q_3(\widetilde{I}P-P\widetilde{I})
\end{eqnarray*}
With the help of the following relations\vspace{-3pt}
\begin{eqnarray*}\widetilde{I}P&=&p_0\widetilde{I}+p_1( {\rm trace \hskip .02in}Z) \widetilde{
I}+p_2\widetilde{I}Z-p_3(Z+\widetilde{I}Z\widetilde{I})
\\
P\widetilde{I}&=&p_0\widetilde{I}+p_1( {\rm trace \hskip .02in}Z) \widetilde{
I}+p_2Z\widetilde{I}+p_3(\widetilde{I}Z\widetilde{I}+Z)
\\
\widetilde{I}P-P\widetilde{I}&=&p_2(\widetilde{I}Z-Z\widetilde{I})-2p_3(Z+ 
\widetilde{I}Z\widetilde{I})
\\
&=&p_2(\widetilde{I}Z-Z\widetilde{I} 
)+2p_3[( {\rm trace \hskip .02in}Z) I-2Z]%\vspace{-15pt}
\end{eqnarray*}
we have \vspace{-3pt}
\begin{eqnarray*}
Q&=&q_0I+ q_1[2p_0+(2p_1+p_2)( {\rm trace \hskip .02in}Z) ]I+\\
&&q_2[p_0I+p_1( {\rm trace \hskip .02in}Z) I+p_2Z+p_3(\widetilde{I}Z-Z 
\widetilde{I})]+ \\
&&q_3\{p_2(\widetilde{I}Z-Z\widetilde{I})+2p_3[( {\rm trace \hskip .02in}Z) I-2Z]\}
\\
 &=&q_0I+2p_0q_1I+p_0q_2I + 
[(2p_1+p_2)q_1+p_1q_2+2p_3q_3]( {\rm trace \hskip .02in}Z) I+ \\
&&(p_2q_2-4p_3q_3)Z+ (p_3q_2+p_2q_3)(\widetilde{I}Z-Z\widetilde{I})\\
&=&r_0I+r_1( {\rm trace \hskip .02in}Z) I+r_2Z+r_3(\widetilde{I}Z-Z\widetilde{I})\vspace{-7pt}
\end{eqnarray*}
and ${\rm col}\{r_0,r_1,r_2,r_3\}=R\ {\rm col}\{q_0,q_1,q_2,q_3\}$ 
where\vspace{-3pt}%\vspace{5pt}%\newpage
\[R=\begin{bmatrix}
1 & 2p_0 & p_0 & 0 \vspace{-5pt}\\ 
0 & 2p_1+p_2 & p_1 & 2p_3 \vspace{-5pt}\\ 
0 & 0 & \,p_2 & -4p_3 \vspace{-5pt}\\ 
0 & 0 & p_3 & \,p_2
\end{bmatrix}\vspace{-3pt}\]
Under the condition that 
${\rm det }\ R =( 2p_1+p_2) ( p_2^2+4p_3^2)\neq 0 $, in the case where $Q=Z$ 
there \vspace{-3pt}

are
${\rm col}\{
q_0 , 
q_1 ,
q_2 ,
q_3
\}=R^{-1}{\rm col}
\{0 ,
0 , 
1 , 
0
\}$ and the inverse function 
 \[Z=q_0I+q_1({\rm trace \hskip .02in}P) I+q_2P+q_3(\widetilde{I}P-P\widetilde{I})\] 
 \begin{Remark}
	Using these expressions one can enter linear constitutive relations for continuous momentless and moment 2-- and 3--dimensional continua which, in the case of their invertibili\-ty, form multiplicative groups. It is an open question about the legality of isotropic maps being unwell--defined (see \cite{Konoplev1999}).%\vspace{-9pt}
\end{Remark} 
\section*{Appendix 5: Multiphase continuum as a Cosserat--Zhilin system}

Following to \cite{Pobedria} let us consider a multiphase continuum, consisting of $n$ components, which may occur between $ m $ chemical reactions (points forming medium assumed to be continuous). During the reaction the proportion of one component decreases, while the other increases. 
Assume that each particulate consists of $ n $ micro--particles (components), so that at each point in space at any given time there are at once all $n$ components, each with a density $\rho_ \alpha (x, t) $ ($ \alpha = 1 , ..., n $). 
Then the total density is defined as $\rho_x = \sum_ {\alpha = 1 } ^ n \rho_ \alpha $.
 
We assume that the center of mass of each micro--particle does not coincide with the center of mass of the particulates. This is the reason to introduce the inertia tensor of particulate components which are recorded, for example in the form of %\vspace{3pt} 
\begin {equation} J_ {ij} = \sum_ {\alpha = 1 } ^ n \rho_ \alpha [(\sum_ {k = 1 } ^ 3 z_k ^ {\alpha} z_k ^ {\alpha}) \delta _ {ij} - z_i ^ {\alpha} z_j ^ {\alpha}] \label{29}\vspace{3pt}   \end{equation}
where $ z_k ^ {\alpha} $ -- coordinates micro--particles $ \alpha $ with respect to particulates, which are assumed to be constant.

Let $ {\it l}_ \alpha = (\sum_ {k = 1 } ^ 3 (z_k ^ {\alpha})
^ {2 } ) ^ {1/ 2} $ be the length of the radius vector micro--particles $ \alpha $ with respect to particulates. Then the first invariant of tensor (\ref{29}) has the form  %\vspace {5pt}%\vspace{5pt}
\[J_x = \sum_ {i = 1 } ^ 3 J_ {ii} = 2 \sum_ {\alpha = 1 } ^ n \rho_ \alpha l_ \alpha ^ 2    \]
 Denoting by $ {v_ \alpha} (x, t) $ velocity of each component, determine the rate of velocity of the particulates as the center of mass of micro--particles%\vspace{-3pt} %\vspace{5pt}
\[ {v_x} = \frac {1 } {\rho_x} \sum_ {\alpha = 1 } ^ n \rho_ \alpha {v_ \alpha}  \]
Introduce the continuity equation for each component of the multiphase medium 
\begin {equation} \frac {\partial \rho_ \alpha} {\partial t} + {\rm div} (\rho_ \alpha {v_ \alpha}) = \gamma_ \alpha \label{30} \end{equation}
where $ \gamma_ \alpha = \sum_ {I = 1 } ^ m \nu_ {\alpha I} J_I $ is called formation of a compound $ \alpha $, the value of $ \nu_ {\alpha I} $ is proportional to the stoichiometric ratio, with which the component $ \alpha $ is included in the $ I $--th chemical reaction; $ J_I $ -- velocity of chemical reactions. 

Summing $ n $ equations (\ref{30}), we arrive at the continuity equation for the density of particulates  \vspace{-9pt}%\vspace{3pt} 
\[\frac {\partial \rho_x} {\partial t} + {\rm div} (\rho_x {v_x}) = \sum_ {\alpha = 1 } ^ n \gamma_ \alpha = \sum_ {\alpha = 1 } ^ n \sum_ {I = 1 } ^ m \nu_ {\alpha I} J_I  \vspace{7pt}\]
Using (\ref{30}), we can write the continuity equation for   \vspace{3pt}
\[\frac {d J_x} {dt} + J_x {\rm div} {v_x} = 2 \sum_ {\alpha = 1 } ^ n \gamma_ \alpha
l_ \alpha ^ 2 \ \ {\text{or}}\ \ \frac {\partial J_x} {\partial t} + {\rm div} (J_x {v_x}) = 2 \sum_ {\alpha = 1 } ^ n \gamma_ \alpha
l_ \alpha ^ 2  \vspace{5pt} \]
Let us introduce%\vspace{-3pt}
\[ {p_x} = \rho_x {v_x}, \quad 
 {q_x} =J_x
 {\mu_x}\vspace{3pt}
 \]
Then using Lemma \ref{Lemma 1} and continuity equations introduced above, from (\ref{3}) we have 6--dimensional differential equation of multiphase continuous medium. It is the result that is given in \cite{Pobedria}. Thus the multiphase system is a Cosserat--Zhilin system of the rather general type.\vspace{-9pt}
 \section*{Conclusion\vspace{-9pt}}
	The author  suggested a new version of rational mechanics and examined the question of their implementation on examples of various  types of Cosserat--Zhilin systems.
	
 He would be highly grateful with whoever would bring any element likely to be able to make progress the development, and thus the comprehension, of the paper. Any comments, reviews,
 critiques, or objections are kindly invited to be sent to him by e--mail. 
%\newpage
\end{document}